\documentclass[11pt]{article}
\usepackage{amsfonts}
\usepackage{color}
\usepackage{graphicx}
\usepackage{authblk}
\usepackage{latexsym,amsmath}
\usepackage{amssymb,array}
\usepackage{enumerate}
\makeatletter \oddsidemargin 0in \evensidemargin 0in \textwidth 16cm
\begin{document}
\newcommand{\la}{\lambda}
\newcommand{\up}{\upsilon}
\newcommand{\Up}{\Upsilon}
\numberwithin{equation}{section}
\newtheorem{d1}{Definition}[section]
\newtheorem{r1}{Remark}[section]
\newtheorem{thm}{Theorem}[section]
\newtheorem{l1}{Lemma}[section]
\newtheorem{ex}{Example}[section]
\newtheorem{cx}{Counterexample}[section]
\newtheorem{cor}{Corollary}[section]
\newcommand{\ts}{\theta}
\newcommand{\Ts}{\theta}
\title{\bf
Comparing Lifetimes of Coherent Systems with Dependent Components Operating in Random Environments}
\author[a]{Nil Kamal Hazra}
\author[b,c]{Maxim Finkelstein\footnote{Corresponding author, email: FinkelM@ufs.ac.za}}
\affil[a]{Department of Mathematics, Indian Institute of Information Technology, Design $\&$ Manufacturing, Kancheepuram, Chennai 600127, India}
\affil[b]{Department of Mathematical Statistics and Actuarial Science, University of the Free State, 339 Bloemfontein 9300, South Africa}
\affil[c]{ITMO University, Saint Petersburg, Russia}
\date{\vspace{-8ex}}
\date{ September 12, 2018}
\maketitle
\begin{abstract}
 We study an impact of a random environment on lifetimes of coherent systems with dependent components. There are two combined sources of this dependence. One results from  the dependence of the components of the coherent system operating in a deterministic environment and the other is due to dependence of components of the system sharing the same random environment. We provide different sets of sufficient conditions for the corresponding stochastic comparisons and consider various scenarios, namely, (i) two different (as a specific case, identical) coherent systems operate under the same random environment; (ii) two coherent systems operate under two different random environments; (iii) one of the coherent systems operates under a random environment, whereas the other under a deterministic one. Some examples are given to illustrate the proposed reasoning.
\end{abstract}
\\\hspace*{3.2 in}Secondary 60E15; 60K10
\section{Introduction and Preliminaries}
Most often, the real-world populations of items are heterogeneous and the corresponding homogeneity can be considered as some approximation. There can be different reasons for heterogeneity. For example, the items can be produced by different manufacturers and then mixed by the user. It can happen with one manufacturer as well, as reliability characteristics of manufactured items change with time depending on many factors (e.g. supplied material,
human factors, the production condition, etc.). Heterogeneity can be also induced by a random environment in which items are operating. This random environment can be modelled, e.g., by the corresponding shock process (see, e.g., Cha and Finkelstein~\cite{cf} and the references therein). However, the simplest and sometimes the most effective way is to model it by the nonnegative environmental random variable $\Theta$ that can affect the distribution of a lifetime $X$ in a baseline deterministic environment (denoted by $F_X(x)$). Thus, conditional on a realization $\Theta=\theta$, the corresponding distribution of $X(\theta)$ is $F_X(x|\theta)$, where the most popular specific models are the scale model $F_X(\theta x)$, the multiplicative and the additive frailty models written as $\theta r_X(x)$ and  $r_X(x)+\theta$, where $r_X(x)$ is the corresponding failure rate assuming that it exists. The most common example of environment is a stress or load under which technical systems are operating (e.g., an electrical load). Some overall climate or nutrition parameters can also describe environment for organisms. An effect of a random environment on various reliability indices was intensively studied in the literature: see, e.g., Finkelstein~\cite{f1}, Petakos and Tsapelas~\cite{pt}, Kenzin and Frostig~\cite{kf}, Nakagawa~\cite{nt}, Persona et al.~\cite{psp}, R$\dot{\rm a}$de~\cite{r}, and the references therein.
\\\hspace*{0.2 in}While describing the following baseline simplified scenarios to be considered in our paper in a much more generality, we will use several basic stochastic orders to be defined for convenience by Definition~\ref{de1} at the end of this section.
\begin{enumerate}
\item [$(i)$] Consider two items (systems) with lifetimes $X(\Theta)$ and $Y(\Theta)$ (we will use later a slightly different notation that is more appropriate for the multicomponent case), operating in the same environment modelled by $\Theta$ and we are interested in stochastic comparisons of these lifetimes. Note that they are dependent via the common environment. Obviously, if we know that its impact is the same on both items (e.g., multiplicative frailty model), then in order, e.g., $X(\Theta)\leq_{st}Y(\Theta)$ to hold, it is sufficient for this inequality to be true for the baseline, deterministic environment, i.e., $X\leq_{st} Y$ as it will hold in each realization of $\Theta$. For a general case, we must just assume this property in each realization, i.e., $X(\theta)\leq_{st}Y(\theta)$, for all $\theta.$
\item [$(ii)$] Let now one item (or two statistically identical items)  operate in two environments $\Theta_1$ and $\Theta_2$ with the corresponding lifetimes $X(\Theta_1)$ and $X(\Theta_2)$. There is a number of simple, meaningful results in the literature for the corresponding comparisons. For example, in Finkelstein~\cite{f2} and Shaked and Shantikumar~\cite{ss}, it is stated that if $\Theta_1\leq_{hr}\Theta_2$, then $X(\Theta_1)\leq_{hr}X(\Theta_2)$ provided that the corresponding failure rate $r(t|\theta)$ is ordered in $\theta$ for all $t>0$.
\item [$(iii)$] The last general introductory setting to be considered is when two items with lifetimes $X(\Theta_1)$ and $Y(\Theta_2)$, are operating in different environments modelled by $\Theta_1$ and $\Theta_2$, respectively. A specific case is when e.g., $\Theta_2$ is degenerate, meaning that the second environment is deterministic.
\end{enumerate}
     The above scenarios are described with respect to comparisons of lifetimes of two items or systems with a \emph{black box} description. However, our paper is dealing with these scenarios for the multicomponent systems, namely, \emph{coherent systems} that satisfy two basic requirements: each component is important for operation of a system and the system lifetime should not decrease if we replace any failed component by a `new' one. This class of systems is rather wide and includes, e.g., the $k$-out-of-$n$ systems as a special case (Barlow and Prochan~\cite{bp}).
     \\\hspace*{0.2 in}Stochastic comparisons for k-out-of-n systems with independent components are extensively studied in Pledger and Proschan~\cite{pp}, Proschan and Sethuraman~\cite{ps}, Balakrishnan and Zhao~\cite{bz}, Hazra et al.~\cite{hkfn}, to name a few. The study of general coherent systems with independent components were considered, e.g., in Esary and Proschan~\cite{ep}, Nanda et al.~\cite{njs}, Kochar et al.~\cite{kms}, Belzunce et al.~\cite{bfrr}, Samaniego~\cite{s}, Hazra and Nanda~\cite{hn}, Samaniego and Navarro~\cite{sn}, Lindqvist et al.~\cite{lsh}.  Stochastic ordering for coherent systems with dependent components is discussed by Navarro and Rubio~\cite{nr}, Navarro et al.~\cite{nass1, nass2, nass3, npd}. An impact of a random environment are studied through multivariate mixture models. Some references to name a few are:  Belzunce et al.~\cite{bmrs}, Misra and Misra~\cite{mm}, Bad$\acute{\rm i}$a et al.~\cite{bsc}, Balakrishnan et al.~\cite{bbh}, and Marshall and Olkin~\cite{mo}.
\\\hspace*{0.2 in}However, to the best of our knowledge, only one paper that we are aware of, is devoted to stochastic comparisons of coherent systems with dependent components operating in different random environments (Amini-Seresht et al.~\cite{azb}). These authors provide some detailed sufficient conditions for the lifetime of a coherent system operating under one (e.g., the more severe) environment to be smaller than that of this system operating under a milder environment. This setting generalizes scenario ($ii$) above to the case of systems with dependent components. It should be noted that there are two `combined' sources of this dependence. One results from the dependence of the components in the coherent system operating in a deterministic environment and the other is due to the dependence of components of the system sharing the same random environment.
\\\hspace*{0.2 in}Inspired by the work of these authors, we present solutions for some \textit{open problems} formulated in Amini-Seresht et al.~\cite{azb}. We also generalize their results and present some new comparisons which consider the settings ($i$) and ($iii$) applied to coherent systems with dependent components. To be more specific, we provide different sets of sufficient conditions for one system to dominate the other one with respect to different stochastic orders, namely, usual stochastic order, hazard rate order, reversed hazard rate order and the likelihood ratio order. It is worth mentioning that the above scenario ($ii$) considered in Amini-Seresht et al.~\cite{azb} for coherent systems can be viewed as a specific case of a more general scenario ($iii$). Moreover, our methodology for obtaining relevant comparisons also differs from that discussed in their paper. Lastly, although the case when e.g., $\Theta_2$ is degenerate, meaning that the second environment is deterministic, is specific for scenario ($iii$), for technical reasons, it is convenient to consider it separately, which is done in Section~\ref{se5}.
\\\hspace*{0.2 in}After this introductory discussion, we can carry on with some basic facts to be used intensively throughout the paper starting with the relevant, formal notation that is more convenient for our multivariate setting. For an absolutely continuous random variable $W$, we denote the probability density function (pdf) by $f_W(\cdot)$, the cumulative distribution function (cdf) by $F_W(\cdot)$, the hazard rate function by $r_W(\cdot)$, the reversed hazard rate function by $\tilde r_W(\cdot)$ and
the survival/reliability
 function by $\bar F_W(\cdot)$.
\\\hspace*{0.3 in}Consider a coherent system with lifetime $\tau\left(\mbox{\boldmath$X$}\right)$ formed by $n$ dependent components with the lifetime vector $\mbox{\boldmath$X$}=(X_1,X_2,\dots,X_n)$.
The dependency among components can be represented by the joint reliability function of $\mbox{\boldmath$X$}$
\begin{eqnarray*}
\bar F_{\mathbf{X}}(x_1,x_2,\dots,x_n)&=&P\left(X_1>x_1,X_2>x_2,\dots,X_n>x_n\right)
\\&=&C\left(\bar F_{X_1}(x_1),\bar F_{X_2}(x_2),\dots,\bar F_{X_n}(x_n)\right),
\end{eqnarray*}
where $C(\cdot,\cdot,\dots,\cdot)$ is a survival copula. In literature, many different types of survival copulas have been studied, for example, Farlie-Gumbel-Morgenstern (FGM) copula, Archimedean copula,
Clayton-Oakes (CO) copula, etc. For more discussion on this, we refer the reader to Nelsen~\cite{n}.
 Based on the above representation, the following fundamental lemma (similar to Theorem 2.1 of Navarro et al.~\cite{nass1}) can be formulated.
\begin{l1}\label{le91}
 Let $\tau\left(\mbox{\boldmath$X$}\right)$ be the lifetime of a coherent system formed by $n$ dependent components with the lifetime vector $\mbox{\boldmath$X$}=(X_1,X_2,\dots,X_n)$. Then the system's reliability function can be written as
 $$\bar F_{\tau\left(\mbox{\boldmath$X$}\right)}(x)=h\left(\bar F_{X_1}(x),\bar F_{X_2}(x),\dots,\bar F_{X_n}(x)\right),$$
 where $h:[0,1]^n\rightarrow [0,1]$, called the domination (or dual distortion) function, depends on the structure function $\phi(\cdot)$ (see Barlow and Proschan~\cite{bp} for definition) and on the survival copula $C$ of $X_1,X_2,\dots,X_n$. Furthermore, $h(\cdot)$ is an increasing continuous function in $[0,1]^n$ such that $h(0,0,\dots,0)=0$ and $h(1,1,\dots,1)=1$. $\hfill\Box$
\end{l1}
\hspace*{0.2 in} An example below illustrates the meaning of this lemma.
\begin{ex}
Consider a coherent system with the lifetime $\tau\left(\mbox{\boldmath$X$}\right)=\min\{X_1,\max\{X_2,X_3\}\}$, where $\mbox{\boldmath$X$}=(X_1,X_2,X_3)$ is described by the FGM Copula (see Nelsen~\cite{n}). Then the minimal path sets (see Barlow and Proschan~\cite{bp}) of this system are given by $P_1=\{1,2\}$ and $P_2=\{1,3\}$. Consequently, its reliability function can be obtained as
\begin{eqnarray*}
\bar F_{\tau\left(\mbox{\boldmath$X$}\right)}(x)&=&P\left(\{X_{P_1}>x\}\cup\{X_{P_2}>x\}\right)
\\&=&P\left(X_{P_1}>x)+P(X_{P_2}>x\right)-P(X_{\{1,2,3\}}>x)
\\&=&\bar F_{\mathbf{X}}(x,x,0)+\bar F_{\mathbf{X}}(x,0,x)-\bar F_{\mathbf{X}}(x,x,x)
\\&=&C\left(\bar F_{X_1}(x),\bar F_{X_2}(x),1\right)+C\left(\bar F_{X_1}(x),1,\bar F_{X_3}(x)\right)+C\left(\bar F_{X_1}(x),\bar F_{X_2}(x),\bar F_{X_3}(x)\right)
\\&=&h\left(\bar F_{X_1}(x),\bar F_{X_2}(x),\bar F_{X_3}(x)\right),
\end{eqnarray*}
where
\begin{eqnarray*}
h(p_1,p_2,p_3)&=&C(p_1,p_2,1)+C(p_1,1,p_2)-C(p_1,p_2,p_3)
\\&=&p_1p_2p_3[1-x(1-p_1)(1-p_2)(1-p_3)], \text{ for }p_i\in(0,1)\text{ and }x\in[-1,1].
\end{eqnarray*}
\end{ex}
\hspace*{0.2 in}Let $\tau\left(\mbox{\boldmath$X$}(\Theta)\right)$ be a random variable representing the lifetime of a coherent system $\tau\left(\mbox{\boldmath$X$}\right)$ that is operating under a random environment modeled by a random variable $\Theta$ with support $\Omega\subseteq [0,\infty).$ For a given environment $\Theta=\theta$, let $\mbox{\boldmath$X$}(\theta)=(X_1(\theta),X_2(\theta),\dots,X_n(\theta))$ be the vector of lifetimes of components, and $h(\cdot)$ be the domination function of $\tau\left(\mbox{\boldmath$X$}(\theta)\right)$. Further, let $\bar F_{X_i}(\cdot|\theta)$, $ F_{X_i}(\cdot|\theta)$ $f_{X_i}(\cdot|\theta)$ $r_{X_i}(\cdot|\theta)$ and $\tilde r_{X_i}(\cdot|\theta)$ be the survival function, the cumulative distribution function, the probability density function, the hazard rate function, and the reversed hazard rate function describing $X_i(\theta)$, respectively, for $i=1,2,\dots,n.$
Then the reliability function describing $\tau\left(\mbox{\boldmath$X$}(\Theta)\right)$ can be expressed as the following mixture
\begin{eqnarray}
\bar F_{\tau\left(\mbox{\boldmath$X$}(\Theta)\right)}(x)&=&P(\tau\left(\mbox{\boldmath$X$}(\Theta)\right)>x)\nonumber
\\&=&\int_\Omega h\left(\bar F_{X_1}(x|\theta),\bar F_{X_2}(x|\theta),\dots,\bar F_{X_n}(x|\theta)\right)dF_\Theta(\theta),\label{df1}
\end{eqnarray}
where the last equality holds due to Lemma~\ref{le91}. Further, its cumulative distribution function is given by
\begin{eqnarray}\label{ef2}
 F_{\tau\left(\mbox{\boldmath$X$}(\Theta)\right)}(x)=\int_\Omega \left[1-h\left(\bar F_{X_1}(x|\theta),\bar F_{X_2}(x|\theta),\dots,\bar F_{X_n}(x|\theta)\right)\right]dF_\Theta(\theta).
\end{eqnarray}
If all $X_i$'s are identical, then \eqref{df1} and \eqref{ef2} reduce to
\begin{eqnarray}\label{ef3}
\bar F_{\tau\left(\mbox{\boldmath$X$}(\Theta)\right)}(x)&=&\int_\Omega h\left(\bar F_{X_1}(x|\theta)\right)dF_\Theta(\theta),
\end{eqnarray}
and
\begin{eqnarray}\label{ef4}
 F_{\tau\left(\mbox{\boldmath$X$}(\Theta)\right)}(x)=\int_\Omega \left[1-h\left(\bar F_{X_1}(x|\theta)\right)\right]dF_\Theta(\theta).
\end{eqnarray}
 \hspace*{0.2 in} Stochastic orders are frequently used as an effective tool to compare the lifetimes of two systems. Numerous stochastic orders are reported in the literature. Each of them has its own merit. For example, the usual stochastic order compares two reliability functions, the hazard rate order compares two hazard/failure rate functions, whereas the reversed hazard rate order compares two reversed hazard rate functions. For exhaustive details see Shaked and Shanthikumar~\cite{ss}. For the sake of completeness, we define below stochastic orders that are used in our paper.
\begin{d1}\label{de1}
Let $X$ and $Y$ be two absolutely continuous random variables with respective supports $(l_X,u_X)$ and $(l_Y,u_Y)$,
where $u_X$ and $u_Y$ may be positive infinity, and $l_X$ and $l_Y$ may be negative infinity.
Then, $X$ is said to be smaller than $Y$ in
\begin{enumerate}
\item [$(a)$] likelihood ratio (lr) order, denoted as $X\leq_{lr}Y$, if
$$\frac{f_Y(t)}{f_X(t)}\;\rm{is\, increasing \, in} \,t\in(l_X,u_X)\cup(l_Y,u_Y);$$
\item hazard rate (hr) order, denoted as $X\leq_{hr}Y$, if $${\bar F_Y(t)}/{\bar F_X(t)}\;\rm{is\,increasing \,in}\, t \in (-\infty,max(u_X,u_Y));$$
\item reversed hazard rate (rhr) order, denoted as $X\leq_{rhr}Y$, if $$ {F_Y(t)}/{ F_X(t)}\;\rm{is\,increasing \,in}\, t \in(min(l_X,l_Y),\infty);$$
 \item usual stochastic (st) order, denoted as $X\leq_{st}Y$, if $$\bar F_X(t)\leq \bar F_Y(t) \text{ for all }t\in~(-\infty,\infty).$$
\end{enumerate}
\end{d1}
Note that the following chain of implications holds among stochastic orders that are discussed above.
$$X\leq_{lr}Y\;\implies X\leq_{hr [rhr]}Y \implies X\leq_{st}Y.$$
 \hspace*{0.2 in}The theory of totally positive functions has vast applications in different areas of approximation theory and related fields. An encyclopedic information on this topic could be found in Karlin~\cite{k}.  Below we give the definitions of TP$_2$ and RR$_2$ functions which will be used in our paper.
\begin{d1}
 Let $\mathcal{X}$ and $\mathcal{Y}$ be two linearly ordered sets. Then, a real-valued function $\kappa(\cdot,\cdot)$ defined on $\mathcal{X}\times\mathcal{Y}$,
 is said to be TP$_2$ (resp. RR$_2$) if
 $$\kappa(x_1,y_1)\kappa(x_2,y_2)\geq(\text{resp. }\leq)\;\kappa(x_1,y_2)\kappa(x_2,y_1),$$
  for all $x_1<x_2$ and $y_1<y_2$.$\hfill \Box$
\end{d1}
\hspace*{0.3 in}Throughout the paper, increasing and decreasing, as usual, mean non-decreasing and non-increasing, respectively. All random variables considered in this paper are assumed to be absolutely continuous and nonnegative. We use bold symbol to represent a vector. Further, we use the acronym `iid' for `independent and identically distributed'.
\\\hspace*{0.3 in}The rest of the paper is organized as follows. In Section~\ref{se2}, we formulate some useful lemmas which are used in proving the main results. In Section~\ref{se3}, we consider two coherent systems that operate in the same random environment. We provide some sufficient conditions for proving that one coherent system dominates the other one with respect to different stochastic orders. In Section~\ref{se4}, we study the same kind of comparisons under the assumption that different coherent systems operate in different random environments. In Section~\ref{se5}, we assume that one of the coherent systems operates in a random environment, whereas the other one in a deterministic environment. The concluding remarks are given in Section~\ref{se6}.
\\\hspace*{0.2 in} To enhance the readability of the paper, all proofs of theorems, wherever given, are deferred to the Appendix.
\section{Useful Lemmas}\label{se2}
In this section we provide some lemmas which will intensively be used in proving the main results. The first lemma describes the TP$_2$/RR$_2$ property of the integral of a function. The proof could be done in the same line as in Lemma 2.1 of Dewan and Khaledi~\cite{dk}.
\begin{l1}\label{le1}
Let $\phi_i(x, \theta)$, $i = 1, 2$, be a nonnegative real valued function on $\mathbb{R} \times \mathbb{X}$, where $\mathbb{R}$ is the set of real numbers, and $\mathbb{X}\subseteq \mathbb{R}$.
Suppose that the following conditions hold.
\begin{itemize}
\item [$(i)$] For $\theta \in \mathbb{X}$, $\phi_2(x, \theta)/\phi_1(x, \theta)$ is [increasing, increasing, decreasing, decreasing, respectively] in $x\in\mathbb{R}$;
\item [$(ii)$] For $x \in \mathbb{R}$, $\phi_2(x, \theta)/\phi_1(x, \theta)$ is [increasing, decreasing, decreasing, increasing, respectively] in $\theta\in \mathbb{X}$;
\item [$(iii)$] Either $\phi_1(x, \theta)$ or $\phi_2(x, \theta)$ is [TP$_2$, RR$_2$, TP$_2$, RR$_2$, respectively] in $(x, \theta)\in \mathbb{R} \times \mathbb{X}$.
\end{itemize}
Then
$$s_i(x) =\int_{X}\phi_i(x, \theta)w(\theta)d\theta \text{ is }
\text{[TP}_2,\text{ TP}_2,\text{ RR}_2,\text{ RR}_2, \text{ respectively ]}\text{ in }(x, i)\in \mathbb{R}\times\{1,2\},$$where $w(\cdot)$ is a continuous function with $\int_{\mathbb{X}} w(\theta)d\theta<\infty$.$\hfill\Box$
\end{l1}
\hspace*{0.2 in}In next four lemmas, we discuss some properties of the reliability functions of the $k$-out-of-$n$ and $l$-out-of-$m$ systems. Lemma~\ref{le2}($i$) and Lemma~\ref{le4}($i$) are obtained in Esary and Proschan~\cite{ep}, whereas Lemma~\ref{le2}($ii$) and Lemma~\ref{le4}($ii$) are obtained in Nanda et al.~\cite{njs}. Further, Lemma~\ref{le2}($iii$), Lemma~\ref{le3}, Lemma~\ref{le5}($i$) and Lemma~\ref{le6} are obtained in Belzunce et al.~\cite{bfrr}. Furthermore, Lemma~\ref{le5}($ii$) and ($iv$) are obtained in Hazra and Nanda~\cite{hn}, whereas Lemma~\ref{le5}($iii$) and ($v$) could be proved in the same line as in Lemmas $5$ and $7$ of Hazra and Nanda~\cite{hn}.
   \begin{l1}\label{le2}
  Let $h_{k:n}(\cdot)$ be the reliability function of the $k$-out-of-$n$ system with iid components.
   Then the following results hold.
 \begin{itemize}
 \item [$(i)$] $ph'_{k:n}(p)/h_{k:n}(p)\;\text{is decreasing in }p\in(0,1)$;
 \item [$(ii)$] $(1-p)h'_{k:n}(p)/(1-h_{k:n}(p))\;\text{is increasing in }p\in(0,1)$;
   \item [$(iii)$] There exists some point $\mu \in (0,1)$ such that
   \begin{itemize}
   \item [$(a)$] $ph''_{k:n}(p)/h'_{k:n}(p)\;\text{is decreasing and positive for all}\;p \in(0,\mu),$ and
  \item [$(b)$] $(1-p)h''_{k:n}(p)/h'_{k:n}(p)\;\text{is decreasing and negative for all}\; p \in(\mu,1).$
  \end{itemize}
  \end{itemize}
  where $\mu=(k-1)/(n-1)$.
   \end{l1}
   \begin{l1}\label{le3}
  Let $h_{k:n}(\cdot)$ and $h_{l:m}(\cdot)$ be the reliability functions of the $k$-out-of-$n$ and the $l$-out-of-$m$ systems with iid components, respectively.
  Then, for $l\leq k$ and $n-k\leq m-l$,
    \begin{itemize}
    \item [$(i)$] $h_{k:n}(p)\leq h_{l:m}(p)$ for all $p\in(0,1)$;
    \item [$(ii)$] $\frac{h_{k:n}(p)}{h_{l:m}(p)} \text{ is increasing }\;p\in(0,1)$;
    \item [$(iii)$] $\frac{1-h_{k:n}(p)}{1-h_{l:m}(p)} \text{ is increasing }\;p\in(0,1)$;
 \item [$(iv)$] $\frac{h_{k:n}'(p)}{h'_{l:m}(p)} \text{ is increasing }\;p\in(0,1).$
 \end{itemize}
 \end{l1}
 \begin{l1} \label{le4}
 Let $h_{k:n}(\cdot)$ and $h_{l:m}(\cdot)$ be the reliability functions of the $k$-out-of-$n$ and of the $l$-out-of-$m$ systems with independent components, respectively.
Then, for $l\leq k$ and $n-k\leq m-l$,
\begin{itemize}
    \item [$(i)$]
 $\sum\limits_{i=1}^n\frac{p_i}{h_{k:n}\left(\mbox{\boldmath$p$}\right)}\frac{\partial h_{k:n}\left(\mbox{\boldmath$p$}\right)}{\partial p_i}$ is decreasing in each $p_i\in(0,1)$, for all $i=1,2,\dots,n$;
   \item [$(ii)$]
 $\sum\limits_{i=1}^n\frac{1-p_i}{1-h_{k:n}\left(\mbox{\boldmath$p$}\right)}\frac{\partial h_{k:n}\left(\mbox{\boldmath$p$}\right)}{\partial p_i}$ is increasing in each $p_i\in(0,1)$, for all $i=1,2,\dots,n$.
 \end{itemize}
\end{l1}
 \begin{l1} \label{le5}
 Let $h_{k:n}(\cdot)$ and $h_{l:m}(\cdot)$ be the reliability functions of the $k$-out-of-$n$ and the $l$-out-of-$m$ systems with independent components, respectively.
Then, for $l\leq k$ and $n-k\leq m-l$,
\begin{itemize}
   \item [$(i)$]
 $h_{k:n}\left(\mbox{\boldmath$p$}\right)\leq h_{l:m}\left(\mbox{\boldmath$p$}\right)$;
 \item [$(ii)$] $\frac{1}{h_{k:n}\left({\bf p}\right)}\frac{\partial h_{k:n}\left(\mbox{\boldmath$p$}\right)}{\partial p_i}\geq \frac{1}{h_{l:m}\left(\mbox{\boldmath$p$}\right)}\frac{\partial h_{l:m}\left({\bf p}\right)}{\partial p_i},$
for all $i=1,2,\dots,\min\{m,n\}$;
\item [$(iii)$] $\frac{1}{1-h_{k:n}\left({\bf p}\right)}\frac{\partial h_{k:n}\left(\mbox{\boldmath$p$}\right)}{\partial p_i}\leq \frac{1}{1-h_{l:m}\left(\mbox{\boldmath$p$}\right)}\frac{\partial h_{l:m}\left({\bf p}\right)}{\partial p_i},$
for all $i=1,2,\dots,\min\{m,n\}$;
    \item [$(iv)$]
 $\sum\limits_{i=1}^n\frac{p_i}{h_{k:n}\left(\mbox{\boldmath$p$}\right)}\frac{\partial h_{k:n}\left(\mbox{\boldmath$p$}\right)}{\partial p_i}\geq \sum\limits_{i=1}^m\frac{p_i}{h_{l:m}\left(\mbox{\boldmath$p$}\right)}\frac{\partial h_{l:m}\left(\mbox{\boldmath$p$}\right)}{\partial p_i}$;
   \item [$(v)$]
 $\sum\limits_{i=1}^n\frac{1-p_i}{1-h_{k:n}\left(\mbox{\boldmath$p$}\right)}\frac{\partial h_{k:n}\left(\mbox{\boldmath$p$}\right)}{\partial p_i}\leq \sum\limits_{i=1}^m\frac{1-p_i}{1-h_{l:m}\left(\mbox{\boldmath$p$}\right)}\frac{\partial h_{l:m}\left(\mbox{\boldmath$p$}\right)}{\partial p_i}.$
 \end{itemize}
\end{l1}
   \begin{l1}\label{le6}
    Let $h_{k:n}(\cdot)$ and $h_{l:m}(\cdot)$ be the reliability functions of the $k$-out-of-$n$ and the $l$-out-of-$m$ systems, respectively.
 Further, let ${\bf Z}=(Z_1,Z_2,\dots,Z_n)$ and ${\bf W}=(W_1,W_2,\dots,W_m)$ be two sets of independent component lifetimes.
  Suppose that, for all $i=1,2,\dots,n$, and $j=1,2,\dots,m$, $Z_i\leq_{lr}W_j$. Then, for $l\leq k$ and $n-k\leq m-l$,
  $$\frac{\partial h_{l:m}(\mbox{\boldmath$q$})}{\partial q_j}\Big /\frac{\partial h_{k:n}(\mbox{\boldmath$p$})}{\partial p_i}\;\text{is increasing in}\;x,$$
  where $p_i=\bar F_{Z_i}(x)$ and $q_j=\bar F_{W_j}(x)$.
 \end{l1}
\section{Two different coherent systems under the same random environment}\label{se3}
In this section, we consider two coherent systems with lifetimes $\tau_1\left(\mbox{\boldmath$X$}(\Theta)\right)$ and $\tau_2\left(\mbox{\boldmath$Y$}(\Theta)\right)$ that operate in the same random environment described by a random variable $\Theta$ with support $\Omega$. For a given (realization) environment $\Theta=\theta$, we denote the domination functions of $\tau_1\left(\mbox{\boldmath$X$}(\theta)\right)$ and  $\tau_2\left(\mbox{\boldmath$Y$}(\theta)\right)$ by $h_1(\cdot)$ and $h_2(\cdot)$, respectively. In what follows, we provide some sufficient conditions for proving that one coherent system dominates the other one with respect to different stochastic orders.
\subsection{Systems with not necessarily identical components}\label{s3}
In this subsection, we consider coherent systems that are formed by not necessarily identical components.
\\\hspace*{0.3 in}In the following theorem, which proof is deferred to the Appendix, we compare two coherent systems with respect to the usual stochastic order.
\begin{thm}\label{t00}
Let $\mbox{\boldmath$X$}=(X_1,X_2,\dots,X_{n})$ and $\mbox{\boldmath$Y$}=(Y_1,Y_2,\dots,Y_{m})$ be two sets of components' lifetimes.
Suppose that the following conditions hold.
\begin{itemize}
\item [$(i)$] $h_1(p_1,p_2,\dots,p_n)\leq h_2(p_1,p_2,\dots,p_m)$;
\item [$(ii)$] $X_i(\theta)\leq_{st}Y_i(\theta)$ for all $i=1,2,\dots,\min\{m,n\}.$
\end{itemize}
Then $\tau_1\left(\mbox{\boldmath$X$}(\Theta)\right)\leq_{st}\tau_2\left(\mbox{\boldmath$Y$}(\Theta)\right)$.$\hfill\Box$
\end{thm}
\hspace*{0.3 in}In the next theorem (see the Appendix for the proof), we show that under some sufficient conditions $\tau_2\left(\mbox{\boldmath$Y$}(\Theta)\right)$ is larger than $\tau_1\left(\mbox{\boldmath$X$}(\Theta)\right)$ with respect to the hazard rate order.
\begin{thm}\label{th22}
Let $\mbox{\boldmath$X$}=(X_1,X_2,\dots,X_{n})$ and $\mbox{\boldmath$Y$}=(Y_1,Y_2,\dots,Y_{m})$ be two sets of components' lifetimes, where $n\geq m$. 
Suppose that $\{(i),(ii),(iii)\}$ or $\{(i),(ii),(iv)\}$ holds.
\begin{itemize}
  \item [$(i)$] $\frac{1}{h_1\left(\mbox{\boldmath$p$}\right)}\frac{\partial h_1\left(\mbox{\boldmath$p$}\right)}{\partial p_i}\geq\frac{1}{{h_2}\left(\mbox{\boldmath$p$}\right)}\frac{\partial {h_2}\left(\mbox{\boldmath$p$}\right)}{\partial p_i}$, for all $i=1,2,\dots,m$;
 \item [$(ii)$] $\frac{p_i}{h_2\left(\mbox{\boldmath$p$}\right)}\frac{\partial h_2\left(\mbox{\boldmath$p$}\right)}{\partial p_i}$ is decreasing in each $p_i$, for all $i=1,2,\dots,m$;
\item [$(iii)$] $X_i(\theta_1)\leq_{hr}X_i(\theta_2),\;\;X_j(\theta)\leq_{hr}Y_j(\theta), \text{ and }Y_j(\theta_2)\leq_{hr}Y_j(\theta_1) \text{ for all }\ts, \theta_1, \theta_2\in \Omega$ such that $\ts_1\leq \ts_2,$ and for all $i=1,2,\dots,n$ and $j=1,2,\dots,m$;
\item [$(iv)$] $X_i(\theta_1)\geq_{hr}X_i(\theta_2),\;\;X_j(\theta)\leq_{hr}Y_j(\theta), \text{ and }Y_j(\theta_2)\geq_{hr}Y_j(\theta_1) \text{ for all }\ts, \theta_1, \theta_2\in \Omega$ such that $\ts_1\leq \ts_2,$ and for all $i=1,2,\dots,n$ and $j=1,2,\dots,m$.
\end{itemize}
Then $\tau_1\left(\mbox{\boldmath$X$}(\Theta)\right)\leq_{hr}\tau_2\left(\mbox{\boldmath$Y$}(\Theta)\right)$.$\hfill\Box$
\end{thm}
\hspace*{0.3 in}The following theorem shows that the same result as in Theorem~\ref{th22} holds for the reversed hazard rate order under some different set of sufficient conditions (see the Appendix for the proof).
\begin{thm}\label{th24}
Let $\mbox{\boldmath$X$}=(X_1,X_2,\dots,X_{n})$ and $\mbox{\boldmath$Y$}=(Y_1,Y_2,\dots,Y_{m})$ be two sets of components' lifetimes, where $m\geq n$. 
Suppose that $\{(i),(ii),(iii)\}$ or $\{(i),(ii),(iv)\}$ holds.
\begin{itemize}
  \item [$(i)$] $\frac{1}{1-h_1\left(\mbox{\boldmath$p$}\right)}\frac{\partial h_1\left(\mbox{\boldmath$p$}\right)}{\partial p_i}\leq\frac{1}{{1-h_2}\left(\mbox{\boldmath$p$}\right)}\frac{\partial {h_2}\left(\mbox{\boldmath$p$}\right)}{\partial p_i}$, for all $i=1,2,\dots,n$;
 \item [$(ii)$] $\frac{1-p_i}{1-h_1\left(\mbox{\boldmath$p$}\right)}\frac{\partial h_1\left(\mbox{\boldmath$p$}\right)}{\partial p_i}$ is increasing in each $p_i$, for all $i=1,2,\dots,n$;
\item [$(iii)$] $X_i(\theta_1)\leq_{rhr}X_i(\theta_2),\;\;X_i(\theta)\leq_{rhr}Y_i(\theta), \text{ and }Y_j(\theta_2)\leq_{rhr}Y_j(\theta_1) \text{ for all }\ts, \theta_1, \theta_2\in \Omega$ such that $\ts_1\leq~\ts_2,$ and for all $i=1,2,\dots,n$ and $j=1,2,\dots,m$;
\item [$(iv)$] $X_i(\theta_1)\geq_{rhr}X_i(\theta_2),\;\;X_i(\theta)\leq_{rhr}Y_i(\theta), \text{ and }Y_j(\theta_2)\geq_{rhr}Y_j(\theta_1) \text{ for all }\ts, \theta_1, \theta_2\in \Omega$ such that $\ts_1\leq~\ts_2,$ and for all $i=1,2,\dots,n$ and $j=1,2,\dots,m$.
\end{itemize}
Then $\tau_1\left(\mbox{\boldmath$X$}(\Theta)\right)\leq_{rhr}\tau_2\left(\mbox{\boldmath$Y$}(\Theta)\right)$.
\end{thm}
\subsection{Systems with iid components}
In this subsection we consider coherent systems of identical components. Obviously, this case has its own value when compared with the general case of non-identical components. The following theorem is analogous to Theorem~\ref{t00}, and the proof also immediately follows from it.
\begin{thm}\label{tr}
Let $\mbox{\boldmath$X$}=(X_1,X_2,\dots,X_{n})$ and $\mbox{\boldmath$Y$}=(Y_1,Y_2,\dots,Y_{m})$ be two sets of components' lifetimes. Assume that $X_i$'s are identical, and that the $Y_j$'s are identical.
Suppose that the following conditions hold.
\begin{itemize}
\item [$(i)$] $h_1(p)\leq h_2(p)$ for all $p\in (0,1)$;
\item [$(ii)$] $X_1(\theta)\leq_{st}Y_1(\theta)$, for all $\theta\in \Omega$.
\end{itemize}
Then $\tau_1\left(\mbox{\boldmath$X$}(\Theta)\right)\leq_{st}\tau_2\left(\mbox{\boldmath$Y$}(\Theta)\right)$.$\hfill\Box$
\end{thm}
\hspace*{0.3 in}In the next theorem we compare two coherent systems with respect to the hazard rate order. The proof could be done in the same line as in Theorem~\ref{th22}.
\begin{thm}\label{t2}
Let $\mbox{\boldmath$X$}=(X_1,X_2,\dots,X_{n})$ and $\mbox{\boldmath$Y$}=(Y_1,Y_2,\dots,Y_{m})$ be two sets of components' lifetimes.
Assume that the $X_i$'s are identical, and that the $Y_j$'s are identical. 
Suppose that $\{(i),(ii),(iii)\}$ or $\{(i),(ii),(iv)\}$ holds.
\begin{itemize}
\item [$(i)$] $h_1(p)/ h_2(p)$ is increasing in $p\in (0,1)$;
\item [$(ii)$] $ph_2'(p)/h_2(p)$ is decreasing in $p\in (0,1)$;
\item [$(iii)$] $X_1(\theta_1)\leq_{hr}X_1(\theta_2)\leq_{hr}Y_1(\theta_2)\leq_{hr}Y_1(\theta_1), \text{ for all }\theta_1, \theta_2\in \Omega\text{ such that }\ts_1\leq \ts_2$;
\item [$(iv)$] $X_1(\theta_2)\leq_{hr}X_1(\theta_1)\leq_{hr}Y_1(\theta_1)\leq_{hr}Y_1(\theta_2), \text{ for all }\theta_1, \theta_2\in \Omega\text{ such that }\ts_1\leq \ts_2.$
\end{itemize}
Then $\tau_1\left(\mbox{\boldmath$X$}(\Theta)\right)\leq_{hr}\tau_2\left(\mbox{\boldmath$Y$}(\Theta)\right)$.$\hfill\Box$
\end{thm}
\hspace*{0.2 in}In the following theorem we show that the same result as in Theorem~\ref{t2} holds for the reversed hazard rate order. The proof is similar to that of Theorem~\ref{th24}.
\begin{thm}\label{t331}
Let $\mbox{\boldmath$X$}=(X_1,X_2,\dots,X_{n})$ and $\mbox{\boldmath$Y$}=(Y_1,Y_2,\dots,Y_{m})$ be two sets of components' lifetimes. Assume that the $X_i$'s are identical, and that the $Y_j$'s are identical.
Suppose that $\{(i),(ii),(iii)\}$ or $\{(i),(ii),(iv)\}$ holds.
\begin{itemize}
\item [$(i)$] $(1-h_1(p))/(1- h_2(p))$ is increasing in $p\in (0,1)$;
\item [$(ii)$] Either $(1-p)h_1'(p)/(1-h_1(p))$ is increasing in $p\in (0,1)$;
\item [$(iii)$] $X_1(\theta_1)\leq_{rhr}X_1(\theta_2)\leq_{rhr}Y_1(\theta_2)\leq_{rhr}Y_1(\theta_1), \text{ for all }\theta_1, \theta_2\in \Omega\text{ such that }\ts_1\leq \ts_2$;
\item [$(iv)$] $X_1(\theta_2)\leq_{rhr}X_1(\theta_1)\leq_{rhr}Y_1(\theta_1)\leq_{rhr}Y_1(\theta_2), \text{ for all }\theta_1, \theta_2\in \Omega\text{ such that }\ts_1\leq \ts_2$.
\end{itemize}
Then $\tau_1\left(\mbox{\boldmath$X$}(\Theta)\right)\leq_{rhr}\tau_2\left(\mbox{\boldmath$Y$}(\Theta)\right)$.
\end{thm}
\section{Different coherent systems under different random environments}\label{se4}
In this section, we consider two coherent systems with lifetimes $\tau_1\left(\mbox{\boldmath$X$}(\Theta_1)\right)$ and $\tau_2\left(\mbox{\boldmath$Y$}(\Theta_2)\right)$, where $\Theta_1$ and $\Theta_2$ are two random variables (with support $\Omega$) that describe two different random environments. For given environments $\Theta_1=\theta$ and $\Theta_2=\theta^*$, we denote the domination functions of $\tau_1\left(\mbox{\boldmath$X$}(\theta)\right)$ and  $\tau_2\left(\mbox{\boldmath$Y$}(\theta^*)\right)$ by $h_1(\cdot)$ and $h_2(\cdot)$, respectively. We will compare $\tau_1\left(\mbox{\boldmath$X$}(\Theta_1)\right)$ and $\tau_2\left(\mbox{\boldmath$Y$}(\Theta_2)\right)$ with respect to different stochastic orders. It should be noted that the results of this section can be considered as generalizations of the corresponding results of Amini-Seresht et al.~\cite{azb} to the case when there are two different coherent systems (in Amini-Seresht et al.~\cite{azb}, the case of one system (or of two identical systems) operating in two environments was discussed).
\subsection{Systems with not necessarily identical components}
In this subsection, we assume that coherent systems are formed by not necessarily identical components.
\\\hspace*{0.2 in}In the following theorem, we show that under a set of sufficient conditions $\tau_2\left(\mbox{\boldmath$Y$}(\Theta_2)\right)$ dominates $\tau_1\left(\mbox{\boldmath$X$}(\Theta_1)\right)$ with respect to the usual stochastic order. The proof follows from Theorem~3.1 of Amini-Seresht et al.~\cite{azb} and our Theorem~\ref{t00}.
\begin{thm}\label{t71}
Let $\mbox{\boldmath$X$}=(X_1,X_2,\dots,X_{n})$ and $\mbox{\boldmath$Y$}=(Y_1,Y_2,\dots,Y_{m})$ be two sets of components' lifetimes.
Suppose that $\{(i),(ii),(iv)\}$ or $\{(i),(iii),(iv)\}$ holds
\begin{itemize}
\item [$(i)$] $h_1(p_1,p_2,\dots,p_n)\leq h_2(p_1,p_2,\dots,p_m)$;
\item [$(ii)$] $X_i(\theta_1)\leq_{st}X_i(\theta_2)$ and $X_j(\theta)\leq_{st}Y_j(\theta)\text{ for all }\ts, \theta_1, \theta_2\in \Omega$ such that $\ts_1\leq \ts_2,$ and for all $i=1,2,\dots,n$ and $j=1,2,\dots,\min\{m,n\};$
\item [$(iii)$] $Y_i(\theta_1)\leq_{st}Y_i(\theta_2)$ and $X_j(\theta)\leq_{st}Y_j(\theta)\text{ for all }\ts, \theta_1, \theta_2\in \Omega$ such that $\ts_1\leq \ts_2,$ and for all $i=1,2,\dots,m$ and $j=1,2,\dots,\min\{m,n\};$
\item [$(iv)$] $\Theta_1\leq_{st} \Theta_2$.
\end{itemize}
Then $\tau_1\left(\mbox{\boldmath$X$}(\Theta_1)\right)\leq_{st}\tau_2\left(\mbox{\boldmath$Y$}(\Theta_2)\right)$.$\hfill\Box$
\end{thm}
\hspace*{0.2 in}The following corollary follows from Theorem~\ref{t71} by using Lemma~\ref{le5}($i$).
\begin{cor}
Let $\mbox{\boldmath$X$}=(X_1,X_2,\dots,X_{n})$ and $\mbox{\boldmath$Y$}=(Y_1,Y_2,\dots,Y_{m})$ be two sets of components' lifetimes. Assume that the $X_i$'s are independent, and that the $Y_j$'s are independent.
Suppose that the set of conditions $\{(ii),(iv)\}$ or $\{(iii),(iv)\}$ in Theorem~\ref{t71} holds. Then
\begin{itemize}
  \item [$(i)$] ${\tau_{k:n}\left(\mbox{\boldmath$X$}(\Theta_1)\right)}\leq_{st}$ ${\tau_{l:n}\left(\mbox{\boldmath$Y$}(\Theta_2)\right)}$ for $l\leq k$;
  \item [$(ii)$] ${\tau_{k:n}\left(\mbox{\boldmath$X$}(\Theta_1)\right)}\leq_{st}$ ${\tau_{k:m}\left(\mbox{\boldmath$Y$}(\Theta_2)\right)}$ for $n\leq m$;
  \item [$(iii)$] ${\tau_{k:n}\left(\mbox{\boldmath$X$}(\Theta_1)\right)}\leq_{st}$ ${\tau_{k-r:n-r}\left(\mbox{\boldmath$Y$}(\Theta_2)\right)}$ for $r\leq k$;
  \item [$(iv)$] ${\tau_{k:n}\left(\mbox{\boldmath$X$}(\Theta_1)\right)}\leq_{st}$ ${\tau_{l:m}\left(\mbox{\boldmath$Y$}(\Theta_2)\right)}$ for $l\leq k$ and $n-k\leq m-l$.$\hfill\Box$
 \end{itemize}
\end{cor}
\hspace*{0.2 in} Now we compare two coherent systems with respect to the hazard rate order. The proof follows from Theorem~3.2 of Amini-Seresht et al.~\cite{azb} and our Theorem~\ref{th22}.
\begin{thm}\label{t72}
Let $\mbox{\boldmath$X$}=(X_1,X_2,\dots,X_{n})$ and $\mbox{\boldmath$Y$}=(Y_1,Y_2,\dots,Y_{m})$ be two sets of components' lifetimes, where $n\geq m$. 
Suppose that $\{(i),(ii),(iii),(v)\}$ or $\{(i),(ii),(iv),(v)\}$ holds.
\begin{itemize}
  \item [$(i)$] $\frac{1}{h_1\left(\mbox{\boldmath$p$}\right)}\frac{\partial h_1\left(\mbox{\boldmath$p$}\right)}{\partial p_i}\geq\frac{1}{{h_2}\left(\mbox{\boldmath$p$}\right)}\frac{\partial {h_2}\left(\mbox{\boldmath$p$}\right)}{\partial p_i}$, for all $i=1,2,\dots,m$;
 \item [$(ii)$] $\frac{p_i}{h_2\left(\mbox{\boldmath$p$}\right)}\frac{\partial h_2\left(\mbox{\boldmath$p$}\right)}{\partial p_i}$ is decreasing in each $p_i$, for all $i=1,2,\dots,m$;
\item [$(iii)$] $X_i(\theta_1)\leq_{hr}X_i(\theta_2),\;\;X_j(\theta)\leq_{hr}Y_j(\theta), \text{ and }Y_j(\theta_2)\leq_{hr}Y_j(\theta_1) \text{ for all }\ts, \theta_1, \theta_2\in \Omega$ such that $\ts_1\leq \ts_2,$ and for all $i=1,2,\dots,n$ and $j=1,2,\dots,m$;
\item [$(iv)$] $X_i(\theta_1)\geq_{hr}X_i(\theta_2),\;\;X_j(\theta)\leq_{hr}Y_j(\theta), \text{ and }Y_j(\theta_2)\geq_{hr}Y_j(\theta_1) \text{ for all }\ts, \theta_1, \theta_2\in \Omega$ such that $\ts_1\leq \ts_2,$ and for all $i=1,2,\dots,n$ and $j=1,2,\dots,m$;
\item [$(v)$] $\Theta_1\leq_{hr} \Theta_2$.
\end{itemize}
Then $\tau_1\left(\mbox{\boldmath$X$}(\Theta_1)\right)\leq_{hr}\tau_2\left(\mbox{\boldmath$Y$}(\Theta_2)\right)$.$\hfill\Box$
\end{thm}
\hspace*{0.3 in}Below we give an example that illustrates conditions ($i$) and ($ii$) of Theorem~\ref{t72}.
\begin{ex}\label{ex1}
 Consider two coherent systems with lifetimes $\tau_1(\mbox{\boldmath$X$})=\min\{X_1,X_2,\dots,X_n\}$ and $\tau_2(\mbox{\boldmath$Y$})=\min\{Y_1,Y_2,\dots,Y_{n-1}\}$, where $n$ is even. Further, let $\{X_1,X_2,\dots,X_{n}\}$ have the Gumbel-Barnett copula given by
 $$C(p_1,p_2,\dots,p_n)=\prod_{i=1}^n p_ie^{-\left(\alpha \prod_{j=1}^n \ln p_j\right)}, \quad \alpha>0,\text{ and }0<p_i<1 \text{ for }i=1,2,\dots,n,$$
 and $\{Y_1,Y_2,\dots,Y_{n-1}\}$ have the Gumbel-Barnett copula given by
  $$C(p_1,p_2,\dots,p_{n-1})=\prod_{i=1}^{n-1} p_ie^{-\left(\alpha \prod_{j=1}^{n-1} \ln p_j\right)}, \quad \alpha>0,\text{ and }0<p_i<1 \text{ for }i=1,2,\dots,n-1.$$
  Then the domination functions of $\tau_1(\mbox{\boldmath$X$})=\min\{X_1,X_2,\dots,X_n\}$ and $\tau_2(\mbox{\boldmath$Y$})=\min\{Y_1,Y_2,\dots,Y_{n-1}\}\}$ are respectively given by
  $$h_1(\mbox{\boldmath$p$})=C(p_1,p_2,\dots,p_n)$$ and $$h_2(\mbox{\boldmath$p$})=C(p_1,p_2,\dots,p_{n-1}).$$Note that
  \begin{eqnarray}\label{eq2}
  \frac{1}{h_1\left(\mbox{\boldmath$p$}\right)}\frac{\partial h_1\left(\mbox{\boldmath$p$}\right)}{\partial p_i}=1-\alpha \prod_{\substack{j= 1,\\ j\neq i}}^n \ln p_j, \text{ for }i=1,2,\dots,n,
  \end{eqnarray}
  and
   \begin{eqnarray}\label{eq3}
  \frac{1}{h_2\left(\mbox{\boldmath$p$}\right)}\frac{\partial h_1\left(\mbox{\boldmath$p$}\right)}{\partial p_i}=1-\alpha \prod_{\substack{j= 1,\\ j\neq i}}^{n-1} \ln p_j, \text{ for }i=1,2,\dots,(n-1).
  \end{eqnarray}
On using \eqref{eq2} and \eqref{eq3},
 $$\frac{1}{h_1\left(\mbox{\boldmath$p$}\right)}\frac{\partial h_1\left(\mbox{\boldmath$p$}\right)}{\partial p_i}\geq\frac{1}{{h_2}\left(\mbox{\boldmath$p$}\right)}\frac{\partial {h_2}\left(\mbox{\boldmath$p$}\right)}{\partial p_i}, \text{ for all } i=1,2,\dots,(n-1),$$
 and
 $$\frac{p_i}{h_2\left(\mbox{\boldmath$p$}\right)}\frac{\partial h_2\left(\mbox{\boldmath$p$}\right)}{\partial p_i} \text{ is decreasing in each }p_i, \text{ for all }i=1,2,\dots,(n-1).$$
$\hfill\Box$
\end{ex}
\hspace*{0.3 in}In the following theorem we provide a set of sufficient conditions for proving that $\tau_2\left(\mbox{\boldmath$Y$}(\Theta_2)\right)$ is larger than $\tau_1\left(\mbox{\boldmath$X$}(\Theta_1)\right)$ with respect to the reversed hazard rate order. The proof follows from Theorem~3.3 of Amini-Seresht et al.~\cite{azb} and Theorem~\ref{th24}.
\begin{thm}\label{t73}
Let $\mbox{\boldmath$X$}=(X_1,X_2,\dots,X_{n})$ and $\mbox{\boldmath$Y$}=(Y_1,Y_2,\dots,Y_{m})$ be two sets of components' lifetimes, where $m\geq n$. 
Suppose that $\{(i),(ii),(iii),(v)\}$ or $\{(i),(ii),(iv),(v)\}$ holds.
\begin{itemize}
  \item [$(i)$] $\frac{1}{1-h_1\left(\mbox{\boldmath$p$}\right)}\frac{\partial h_1\left(\mbox{\boldmath$p$}\right)}{\partial p_i}\leq\frac{1}{{1-h_2}\left(\mbox{\boldmath$p$}\right)}\frac{\partial {h_2}\left(\mbox{\boldmath$p$}\right)}{\partial p_i}$, for all $i=1,2,\dots,n$;
 \item [$(ii)$] $\frac{1-p_i}{1-h_1\left(\mbox{\boldmath$p$}\right)}\frac{\partial h_1\left(\mbox{\boldmath$p$}\right)}{\partial p_i}$ is increasing in each $p_i$, for all $i=1,2,\dots,n$;
\item [$(iii)$] $X_i(\theta_1)\leq_{rhr}X_i(\theta_2),\;\;X_i(\theta)\leq_{rhr}Y_i(\theta), \text{ and }Y_j(\theta_2)\leq_{rhr}Y_j(\theta_1) \text{ for all }\ts, \theta_1, \theta_2\in \Omega$ such that $\ts_1\leq~\ts_2,$ and for all $i=1,2,\dots,n$ and $j=1,2,\dots,m$;
\item [$(iv)$] $X_i(\theta_1)\geq_{rhr}X_i(\theta_2),\;\;X_i(\theta)\leq_{rhr}Y_i(\theta), \text{ and }Y_j(\theta_2)\geq_{rhr}Y_j(\theta_1) \text{ for all }\ts, \theta_1, \theta_2\in \Omega$ such that $\ts_1\leq~\ts_2,$ and for all $i=1,2,\dots,n$ and $j=1,2,\dots,m$;
\item [$(v)$] $\Theta_1\leq_{rhr} \Theta_2$.
\end{itemize}
Then $\tau_1\left(\mbox{\boldmath$X$}(\Theta_1)\right)\leq_{rhr}\tau_2\left(\mbox{\boldmath$Y$}(\Theta_2)\right)$.
\end{thm}
\subsection{Systems with iid components}
In this subsection we compare two coherent systems that are formed by iid components. Obviously, it has its own value when compared with the general case of non-identical components.
\\\hspace*{0.2 in}We show that under a set of sufficient conditions $\tau_2\left(\mbox{\boldmath$Y$}(\Theta_2)\right)$ is larger than $\tau_1\left(\mbox{\boldmath$X$}(\Theta_1)\right)$ with respect to the usual stochastic order. The proof follows from Theorem~3.1 of Amini-Seresht et al.~\cite{azb} and Theorem~\ref{tr}.
\begin{thm}\label{t61}
Let $\mbox{\boldmath$X$}=(X_1,X_2,\dots,X_{n})$ and $\mbox{\boldmath$Y$}=(Y_1,Y_2,\dots,Y_{m})$ be two sets of components' lifetimes. Assume that $X_i$'s are identical, and that the $Y_j$'s are identical.
Suppose that $\{(i),(ii),(iv)\}$ or $\{(i),(iii),(iv)\}$ holds.
\begin{itemize}
\item [$(i)$] $h_1(p)\leq h_2(p)$ for all $p\in (0,1)$;
\item [$(ii)$] $X_1(\theta_1)\leq_{st}X_1(\theta_2)$ and $X_1(\theta)\leq_{st}Y_1(\theta), \text{ for all }\theta,\theta_1, \theta_2\in \Omega\text{ such that }\ts_1\leq \ts_2$;
\item [$(iii)$] $Y_1(\theta_1)\leq_{st}Y_1(\theta_2)$ and $X_1(\theta)\leq_{st}Y_1(\theta), \text{ for all }\theta,\theta_1, \theta_2\in \Omega\text{ such that }\ts_1\leq \ts_2$;
\item [$(iv)$] $\Theta_1\leq_{st} \Theta_2$.
\end{itemize}
Then $\tau_1\left(\mbox{\boldmath$X$}(\Theta_1)\right)\leq_{st}\tau_2\left(\mbox{\boldmath$Y$}(\Theta_2)\right)$.$\hfill\Box$
\end{thm}
\hspace*{0.3 in}The next corollary immediately follows from Theorem~\ref{t61} by using Lemma~\ref{le3}.
\begin{cor}
Let $\mbox{\boldmath$X$}=(X_1,X_2,\dots,X_{n})$ and $\mbox{\boldmath$Y$}=(Y_1,Y_2,\dots,Y_{m})$ be two sets of components' lifetimes. Assume that $X_i$'s are iid, and that the $Y_j$'s are iid.
  Suppose that the set of conditions $\{(ii),(iv)\}$ or $\{(iii),(iv)\}$ in Theorem~\ref{t61} holds. Then
\begin{itemize}
  \item [$(i)$] ${\tau_{k:n}\left(\mbox{\boldmath$X$}(\Theta_1)\right)}\leq_{st}$ ${\tau_{l:n}\left(\mbox{\boldmath$Y$}(\Theta_2)\right)}$ for $l\leq k$;
  \item [$(ii)$] ${\tau_{k:n}\left(\mbox{\boldmath$X$}(\Theta_1)\right)}\leq_{st}$ ${\tau_{k:m}\left(\mbox{\boldmath$Y$}(\Theta_2)\right)}$ for $n\leq m$;
  \item [$(iii)$] ${\tau_{k:n}\left(\mbox{\boldmath$X$}(\Theta_1)\right)}\leq_{st}$ ${\tau_{k-r:n-r}\left(\mbox{\boldmath$Y$}(\Theta_2)\right)}$ for $r\leq k$;
  \item [$(iv)$] ${\tau_{k:n}\left(\mbox{\boldmath$X$}(\Theta_1)\right)}\leq_{st}$ ${\tau_{l:m}\left(\mbox{\boldmath$Y$}(\Theta_2)\right)}$ for $l\leq k$ and $n-k\leq m-l$.$\hfill\Box$
 \end{itemize}
\end{cor}
\hspace*{0.2 in}The next theorem discusses the same result as in the above theorem but with respect to the hazard rate order. The proof follows from corollary~3.1 of Amini-Seresht et al.~\cite{azb} and Theorem~\ref{t2}.
\begin{thm}\label{t62}
Let $\mbox{\boldmath$X$}=(X_1,X_2,\dots,X_{n})$ and $\mbox{\boldmath$Y$}=(Y_1,Y_2,\dots,Y_{m})$ be two sets of components' lifetimes. Assume that $X_i$'s are identical, and that the $Y_j$'s are identical.
Suppose that $\{(i),(ii),(iii),(v)\}$ or $\{(i),(ii),(iv),(v)\}$ holds.
\begin{itemize}
\item [$(i)$] $h_1(p)/ h_2(p)$ is increasing in $p\in (0,1)$;
\item [$(ii)$] $ph_2'(p)/h_2(p)$ is decreasing in $p\in (0,1)$;
\item [$(iii)$] $X_1(\theta_1)\leq_{hr}X_1(\theta_2)\leq_{hr}Y_1(\theta_2)\leq_{hr}Y_1(\theta_1), \text{ for all }\theta_1, \theta_2\in \Omega\text{ such that }\ts_1\leq \ts_2$;
\item [$(iv)$] $X_1(\theta_2)\leq_{hr}X_1(\theta_1)\leq_{hr}Y_1(\theta_1)\leq_{hr}Y_1(\theta_2), \text{ for all }\theta_1, \theta_2\in \Omega\text{ such that }\ts_1\leq \ts_2$;
\item [$(v)$] $\Theta_1\leq_{hr} \Theta_2$.
\end{itemize}
Then $\tau_1\left(\mbox{\boldmath$X$}(\Theta_1)\right)\leq_{hr}\tau_2\left(\mbox{\boldmath$Y$}(\Theta_2)\right)$.$\hfill\Box$
\end{thm}
\hspace*{0.2 in}The following corollary follows from Theorem~\ref{t62} by using Lemma~\ref{le2}($i$) and Lemma~\ref{le3}($ii$).
\begin{cor}
 Let $\mbox{\boldmath$X$}=(X_1,X_2,\dots,X_{n})$ and $\mbox{\boldmath$Y$}=(Y_1,Y_2,\dots,Y_{m})$ be two sets of components' lifetimes. Assume that the $X_i$'s are iid, and that the $Y_j$'s are iid.
  Suppose that the set of conditions $\{(iii), (v)\}$ or $\{(iv),(v)\}$ in Theorem~\ref{t62} holds. Then
 \begin{itemize}
  \item [$(i)$] ${\tau_{k:n}\left(\mbox{\boldmath$X$}(\Theta_1)\right)}\leq_{hr}$ ${\tau_{l:n}\left(\mbox{\boldmath$Y$}(\Theta_2)\right)}$ for $l\leq k$;
  \item [$(ii)$] ${\tau_{k:n}\left(\mbox{\boldmath$X$}(\Theta_1)\right)}\leq_{hr}$ ${\tau_{k:m}\left(\mbox{\boldmath$Y$}(\Theta_2)\right)}$ for $n\leq m$;
  \item [$(iii)$] ${\tau_{k:n}\left(\mbox{\boldmath$X$}(\Theta_1)\right)}\leq_{hr}$ ${\tau_{k-r:n-r}\left(\mbox{\boldmath$Y$}(\Theta_2)\right)}$ for $r\leq k$;
  \item [$(iv)$] ${\tau_{k:n}\left(\mbox{\boldmath$X$}(\Theta_1)\right)}\leq_{hr}$ ${\tau_{l:m}\left(\mbox{\boldmath$Y$}(\Theta_2)\right)}$ for $l\leq k$ and $n-k\leq m-l$.$\hfill\Box$
 \end{itemize}
\end{cor}
\hspace*{0.2 in}The example to follow illustrates conditions ($i$) and ($ii$) of Theorem~\ref{t62}.
\begin{ex}\label{ex1}
 Consider two coherent systems with lifetimes $\tau_1(\mbox{\boldmath$X$})=\min\{X_1,X_2,X_3\}$ and $\tau_2(\mbox{\boldmath$Y$})=\min\{Y_1,\max\{Y_2,Y_3\}\}$, where both $\{X_1,X_2,X_3\}$ and $\{Y_1,Y_2,Y_3\}$ are homogeneous and have the same FGM copula given by
 $$C(p_1,p_2,p_3)=p_1p_2p_3(1+x(1-p_1)(1-p_2)(1-p_3)),\quad\text{for}\;x \in[-1,1].$$
 Then the domination functions of $\tau_1(\mbox{\boldmath$X$})=\min\{X_1,X_2,X_3\}$ and $\tau_2(\mbox{\boldmath$Y$})=\min\{Y_1,\max\{Y_2,Y_3\}\}$ are given by
\begin{eqnarray*}
 h_1(p)=C(p,p,p)=p^3(1+x(1-p)^3),
\end{eqnarray*}
and
$$h_2(p)=2C(p,p,1)-C(p,p,p)=2p^2-p^3-xp^3(1-p)^3,$$
respectively. Further, it could be verified that, $\text{for all }x\in [-1,1],$
$$\frac{h_1(p)}{h_2(p)}=\frac{p^3(1+x(1-p)^3)}{2p^2-p^3(1+x(1-p)^3)}\text{ is increasing in }p\in(0,1),$$ and
$$p\frac{h_2'(p)}{h_2(p)}=\frac{4-3(1+x)p+12xp^2-15xp^3+6xp^4}{2-(1+x)p+3xp^2-3xp^3+xp^4}\text{ is decreasing in }p\in(0,1),$$
and hence our claim is proved.
$\hfill\Box$
\end{ex}
\hspace*{0.2 in}In the next theorem, we prove the similar result as in Theorem~\ref{t62}, but for the reversed hazard rate order. The proof follows from corollary~3.2 of Amini-Seresht et al.~\cite{azb} and Theorem~\ref{t331}.
\begin{thm}\label{t63}
Let $\mbox{\boldmath$X$}=(X_1,X_2,\dots,X_{n})$ and $\mbox{\boldmath$Y$}=(Y_1,Y_2,\dots,Y_{m})$ be two sets of components' lifetimes. Assume that the $X_i$'s are identical, and that the $Y_j$'s are identical.
Suppose that $\{(i),(ii),(iii),(v)\}$ or $\{(i),(ii),(iv),(v)\}$ holds.
\begin{itemize}
\item [$(i)$] $(1-h_1(p))/(1- h_2(p))$ is increasing in $p\in (0,1)$;
\item [$(ii)$] $(1-p)h_1'(p)/(1-h_1(p))$ is increasing in $p\in (0,1)$;
\item [$(iii)$] $X_1(\theta_1)\leq_{rhr}X_1(\theta_2)\leq_{rhr}Y_1(\theta_2)\leq_{rhr}Y_1(\theta_1), \text{ for all }\theta_1, \theta_2\in \Omega\text{ such that }\ts_1\leq \ts_2$;
\item [$(iv)$] $X_1(\theta_2)\leq_{rhr}X_1(\theta_1)\leq_{rhr}Y_1(\theta_1)\leq_{rhr}Y_1(\theta_2), \text{ for all }\theta_1, \theta_2\in \Omega\text{ such that }\ts_1\leq \ts_2$;
\item [$(v)$] $\Theta_1\leq_{rhr} \Theta_2$.
\end{itemize}
Then $\tau_1\left(\mbox{\boldmath$X$}(\Theta_1)\right)\leq_{rhr}\tau_2\left(\mbox{\boldmath$Y$}(\Theta_2)\right)$.$\hfill\Box$
\end{thm}
\hspace*{0.2 in}The next corollary follows from Theorem~\ref{t63} by using Lemma~\ref{le2}($ii$) and Lemma~\ref{le3}($iii$).
\begin{cor}
 Let $\mbox{\boldmath$X$}=(X_1,X_2,\dots,X_{n})$ and $\mbox{\boldmath$Y$}=(Y_1,Y_2,\dots,Y_{m})$ be two sets of components' lifetimes. Assume that $X_i$'s are iid, and that the $Y_j$'s are iid.
 Suppose that the set of conditions $\{(iii), (v)\}$ or $\{(iv),(v)\}$ in Theorem~\ref{t63} holds. Then
 \begin{itemize}
  \item [$(i)$] ${\tau_{k:n}\left(\mbox{\boldmath$X$}(\Theta_1)\right)}\leq_{rhr}$ ${\tau_{l:n}\left(\mbox{\boldmath$Y$}(\Theta_2)\right)}$ for $l\leq k$;
  \item [$(ii)$] ${\tau_{k:n}\left(\mbox{\boldmath$X$}(\Theta_1)\right)}\leq_{rhr}$ ${\tau_{k:m}\left(\mbox{\boldmath$Y$}(\Theta_2)\right)}$ for $n\leq m$;
  \item [$(iii)$] ${\tau_{k:n}\left(\mbox{\boldmath$X$}(\Theta_1)\right)}\leq_{rhr}$ ${\tau_{k-r:n-r}\left(\mbox{\boldmath$Y$}(\Theta_2)\right)}$ for $r\leq k$;
  \item [$(iv)$] ${\tau_{k:n}\left(\mbox{\boldmath$X$}(\Theta_1)\right)}\leq_{rhr}$ ${\tau_{l:m}\left(\mbox{\boldmath$Y$}(\Theta_2)\right)}$ for $l\leq k$ and $n-k\leq m-l$.
 \end{itemize}
\end{cor}
\section{One of the coherent systems under a random environment}\label{se5}
In this section, as previously, we compare two coherent systems with respect to different stochastic orders. However, we assume that one of them operates in a random environment, whereas the other one operates in a deterministic environment. As was already mentioned, although this case can be viewed as the special case of the discussion in the previous section, technically it is more convenient to consider it independently. Let $\tau_1\left(\mbox{\boldmath$X$}(\Theta)\right)$ be the lifetime of a coherent system that operates in a random environment modeled by a random variable $\Theta$ with support $\Omega$. Further, let $\tau_2\left(\mbox{\boldmath$Y$}\right)$ be the lifetime of the other coherent system that operates in some baseline, deterministic environment.  For a given environment $\Theta=\theta$, we denote the domination function of $\tau_1\left(\mbox{\boldmath$X$}(\theta)\right)$ by $h_1(\cdot)$. Further, we denote the domination function of $\tau_2\left(\mbox{\boldmath$Y$}\right)$ by $h_2(\cdot)$.
\subsection{Systems with not necessarily identical components}\label{s51}
\hspace*{0.3 in}In the following theorem we show that under a set of sufficient conditions $\tau_1\left(\mbox{\boldmath$X$}(\Theta)\right)$ dominates $\tau_2\left(\mbox{\boldmath$Y$}\right)$ with respect to the usual stochastic order.
The proof follows in the same line as in Theorem~\ref{t00}, and hence omitted.
\begin{thm}\label{t81}
Let $\mbox{\boldmath$X$}=(X_1,X_2,\dots,X_{n})$ and $\mbox{\boldmath$Y$}=(Y_1,Y_2,\dots,Y_{m})$ be two sets of components' lifetimes.
Suppose that the following conditions hold.
\begin{itemize}
\item [$(i)$] $h_1(p_1,p_2,\dots,p_n)\leq h_2(p_1,p_2,\dots,p_m)$;
\item [$(ii)$] $X_i(\theta)\leq_{st}Y_i \text{ for all }\theta\in \Omega,$ and for all $i=1,2,\dots,\min\{m,n\}.$
\end{itemize}
Then $\tau_1\left(\mbox{\boldmath$X$}(\Theta)\right)\leq_{st}\tau_2\left(\mbox{\boldmath$Y$}\right)$.$\hfill\Box$
\end{thm}
\hspace*{0.2 in}The next corollary immediately follows from Theorem~\ref{t81} by using Lemma~\ref{le5}($i$).
\begin{cor}
 Let $\mbox{\boldmath$X$}=(X_1,X_2,\dots,X_{n})$ and $\mbox{\boldmath$Y$}=(Y_1,Y_2,\dots,Y_{m})$ be two sets of components' lifetimes. Assume that the $X_i$'s are independent, and that the $Y_j$'s are independent.
 Suppose that condition ($ii$) in Theorem~\ref{t81} holds. Then
 \begin{itemize}
  \item [$(i)$] ${\tau_{k:n}\left(\mbox{\boldmath$X$}(\Theta)\right)}\leq_{st}$ ${\tau_{l:n}\left(\mbox{\boldmath$Y$}\right)}$ for $l\leq k$;
  \item [$(ii)$] ${\tau_{k:n}\left(\mbox{\boldmath$X$}(\Theta)\right)}\leq_{st}$ ${\tau_{k:m}\left(\mbox{\boldmath$Y$}\right)}$ for $n\leq m$;
  \item [$(iii)$] ${\tau_{k:n}\left(\mbox{\boldmath$X$}(\Theta)\right)}\leq_{st}$ ${\tau_{k-r:n-r}\left(\mbox{\boldmath$Y$}\right)}$ for $r\leq k$;
  \item [$(iv)$] ${\tau_{k:n}\left(\mbox{\boldmath$X$}(\Theta)\right)}\leq_{st}$ ${\tau_{l:m}\left(\mbox{\boldmath$Y$}\right)}$ for $l\leq k$ and $n-k\leq m-l$.$\hfill\Box$
 \end{itemize}
\end{cor}
\hspace*{0.2 in} The theorem to follow, provides some sufficient conditions for proving that $\tau_1\left(\mbox{\boldmath$X$}(\Theta)\right)$ is smaller than $\tau_2\left(\mbox{\boldmath$Y$}\right)$ with respect to the hazard rate order. See the Appendix for the proof.
\begin{thm}\label{t82}
Let $\mbox{\boldmath$X$}=(X_1,X_2,\dots,X_{n})$ and $\mbox{\boldmath$Y$}=(Y_1,Y_2,\dots,Y_{m})$ be two sets of components' lifetimes.
Suppose that $\{(i),(ii),(iv)\}$ or $\{(i),(iii),(iv)\}$ holds.
\begin{itemize}
  \item [$(i)$] $\sum\limits_{i=1}^n\frac{p_i}{h_1\left(\mbox{\boldmath$p$}\right)}\frac{\partial h_1\left(\mbox{\boldmath$p$}\right)}{\partial p_i}\geq \sum\limits_{i=1}^m\frac{p_i}{{h_2}\left(\mbox{\boldmath$p$}\right)}\frac{\partial {h_2}\left(\mbox{\boldmath$p$}\right)}{\partial p_i}$;
 \item [$(ii)$] $\sum\limits_{i=1}^n\frac{p_i}{h_1\left(\mbox{\boldmath$p$}\right)}\frac{\partial h_1\left(\mbox{\boldmath$p$}\right)}{\partial p_i}$  is decreasing in each $p_i$, $i=1,2,\dots,n$;
 \item [$(iii)$] $\sum\limits_{i=1}^n\frac{p_i}{h_2\left(\mbox{\boldmath$p$}\right)}\frac{\partial h_2\left(\mbox{\boldmath$p$}\right)}{\partial p_i}$  is decreasing in each $p_i$, $i=1,2,\dots,m$;
\item [$(iv)$] $X_i(\theta)\leq_{hr}Y_j \text{ for all }\theta\in \Omega,$ and for all $i=1,2,\dots,n$ and $j=1,2,\dots,m.$
\end{itemize}
Then $\tau_1\left(\mbox{\boldmath$X$}(\Theta)\right)\leq_{hr}\tau_2\left(\mbox{\boldmath$Y$}\right)$.$\hfill\Box$
\end{thm}
\hspace*{0.3 in}On using Lemma~\ref{le4}($i$) and Lemma~\ref{le5}($iv$),  from Theorem~\ref{t82}:
\begin{cor}
 Let $\mbox{\boldmath$X$}=(X_1,X_2,\dots,X_{n})$ and $\mbox{\boldmath$Y$}=(Y_1,Y_2,\dots,Y_{m})$ be two sets of components' lifetimes. Assume that the $X_i$'s are independent, and that the $Y_j$'s are independent.
 Suppose that condition ($iv$) of Theorem~\ref{t82} holds. Then
 \begin{itemize}
  \item [$(i)$] ${\tau_{k:n}\left(\mbox{\boldmath$X$}(\Theta)\right)}\leq_{hr}$ ${\tau_{l:n}\left(\mbox{\boldmath$Y$}\right)}$ for $l\leq k$;
  \item [$(ii)$] ${\tau_{k:n}\left(\mbox{\boldmath$X$}(\Theta)\right)}\leq_{hr}$ ${\tau_{k:m}\left(\mbox{\boldmath$Y$}\right)}$ for $n\leq m$;
  \item [$(iii)$] ${\tau_{k:n}\left(\mbox{\boldmath$X$}(\Theta)\right)}\leq_{hr}$ ${\tau_{k-r:n-r}\left(\mbox{\boldmath$Y$}\right)}$ for $r\leq k$;
  \item [$(iv)$] ${\tau_{k:n}\left(\mbox{\boldmath$X$}(\Theta)\right)}\leq_{hr}$ ${\tau_{l:m}\left(\mbox{\boldmath$Y$}\right)}$ for $l\leq k$ and $n-k\leq m-l$.$\hfill\Box$
 \end{itemize}
\end{cor}
\hspace*{0.2 in}Even though the result given in the following theorem is the same as in Theorem~\ref{t82}, the set of sufficient conditions used here is different from the previous one. The proof could be performed in the same line as in Theorem~\ref{t82} and hence omitted.
\begin{thm}\label{t882}
Let $\mbox{\boldmath$X$}=(X_1,X_2,\dots,X_{n})$ and $\mbox{\boldmath$Y$}=(Y_1,Y_2,\dots,Y_{m})$ be two sets of components' lifetimes with $n\geq m$. 
Suppose that the following conditions hold.
\begin{itemize}
  \item [$(i)$] $\frac{p_i}{h_1\left(\mbox{\boldmath$p$}\right)}\frac{\partial h_1\left(\mbox{\boldmath$p$}\right)}{\partial p_i}\geq \frac{p_i}{{h_2}\left(\mbox{\boldmath$p$}\right)}\frac{\partial {h_2}\left(\mbox{\boldmath$p$}\right)}{\partial p_i}$, for all $i=1,2,\dots,m$;
 \item [$(ii)$] $\frac{p_i}{h_1\left(\mbox{\boldmath$p$}\right)}\frac{\partial h_1\left(\mbox{\boldmath$p$}\right)}{\partial p_i}$ or $\frac{p_i}{h_2\left(\mbox{\boldmath$p$}\right)}\frac{\partial h_2\left(\mbox{\boldmath$p$}\right)}{\partial p_i}$ is decreasing in each $p_i$, $i=1,2,\dots,m$;
\item [$(iii)$] $X_i(\theta)\leq_{hr}Y_i \text{ for all }\theta\in \Omega,$ and for all $i=1,2,\dots,m.$
\end{itemize}
Then $\tau_1\left(\mbox{\boldmath$X$}(\Theta)\right)\leq_{hr}\tau_2\left(\mbox{\boldmath$Y$}\right)$.$\hfill\Box$
\end{thm}
\hspace*{0.3 in} Now we discuss the corresponding result for the reversed hazard rate order. The proof is deferred to the Appendix.
\begin{thm}\label{t83}
Let $\mbox{\boldmath$X$}=(X_1,X_2,\dots,X_{n})$ and $\mbox{\boldmath$Y$}=(Y_1,Y_2,\dots,Y_{m})$ be two sets of components' lifetimes.
Suppose that $\{(i),(ii),(iv)\}$ or $\{(i),(iii),(iv)\}$ holds.
\begin{itemize}
  \item [$(i)$] $\sum\limits_{i=1}^n\frac{1-p_i}{1-h_1\left(\mbox{\boldmath$p$}\right)}\frac{\partial h_1\left(\mbox{\boldmath$p$}\right)}{\partial p_i}\leq \sum\limits_{i=1}^m\frac{1-p_i}{1-{h_2}\left(\mbox{\boldmath$p$}\right)}\frac{\partial {h_2}\left(\mbox{\boldmath$p$}\right)}{\partial p_i}$;
 \item [$(ii)$] $\sum\limits_{i=1}^n\frac{1-p_i}{1-h_1\left(\mbox{\boldmath$p$}\right)}\frac{\partial h_1\left(\mbox{\boldmath$p$}\right)}{\partial p_i}$  is increasing in each $p_i$, $i=1,2,\dots,n$;
  \item [$(iii)$] $\sum\limits_{i=1}^n\frac{1-p_i}{1-h_2\left(\mbox{\boldmath$p$}\right)}\frac{\partial h_2\left(\mbox{\boldmath$p$}\right)}{\partial p_i}$  is increasing in each $p_i$, $i=1,2,\dots,m$;
\item [$(iv)$] $X_i(\theta)\leq_{rhr}Y_j \text{ for all }\theta\in \Omega,$ and for all $i=1,2,\dots,n$ and $j=1,2,\dots,m.$
\end{itemize}
Then $\tau_1\left(\mbox{\boldmath$X$}(\Theta)\right)\leq_{rhr}\tau_2\left(\mbox{\boldmath$Y$}\right)$.$\hfill\Box$
\end{thm}
\hspace*{0.3 in} This corollary follows from Theorem~\ref{t83} by using Lemma~\ref{le4}($ii$) and Lemma~\ref{le5}($v$).
\begin{cor}
 Let $\mbox{\boldmath$X$}=(X_1,X_2,\dots,X_{n})$ and $\mbox{\boldmath$Y$}=(Y_1,Y_2,\dots,Y_{m})$ be two sets of components' lifetimes. Assume that the $X_i$'s are independent, and that the $Y_j$'s are independent.
  Suppose that condition ($iv$) in Theorem~\ref{t83} holds. Then
 \begin{itemize}
  \item [$(i)$] ${\tau_{k:n}\left(\mbox{\boldmath$X$}(\Theta)\right)}\leq_{rhr}$ ${\tau_{l:n}\left(\mbox{\boldmath$Y$}\right)}$ for $l\leq k$;
  \item [$(ii)$] ${\tau_{k:n}\left(\mbox{\boldmath$X$}(\Theta)\right)}\leq_{rhr}$ ${\tau_{k:m}\left(\mbox{\boldmath$Y$}\right)}$ for $n\leq m$;
  \item [$(iii)$] ${\tau_{k:n}\left(\mbox{\boldmath$X$}(\Theta)\right)}\leq_{rhr}$ ${\tau_{k-r:n-r}\left(\mbox{\boldmath$Y$}\right)}$ for $r\leq k$;
  \item [$(iv)$] ${\tau_{k:n}\left(\mbox{\boldmath$X$}(\Theta)\right)}\leq_{rhr}$ ${\tau_{l:m}\left(\mbox{\boldmath$Y$}\right)}$ for $l\leq k$ and $n-k\leq m-l$.$\hfill\Box$
 \end{itemize}
\end{cor}
\hspace*{0.2 in}In the next theorem we show that the same result as in Theorem~\ref{t83} holds under a different set of sufficient conditions. The proof is similar to the previous theorem, and hence omitted.
\begin{thm}\label{t883}
Let $\mbox{\boldmath$X$}=(X_1,X_2,\dots,X_{n})$ and $\mbox{\boldmath$Y$}=(Y_1,Y_2,\dots,Y_{m})$ be two sets of components' lifetimes with $m\geq n$.
Suppose that the following conditions hold.
\begin{itemize}
  \item [$(i)$] $\frac{1-p_i}{1-h_1\left(\mbox{\boldmath$p$}\right)}\frac{\partial h_1\left(\mbox{\boldmath$p$}\right)}{\partial p_i}\leq \frac{1-p_i}{1-{h_2}\left(\mbox{\boldmath$p$}\right)}\frac{\partial {h_2}\left(\mbox{\boldmath$p$}\right)}{\partial p_i}$;
 \item [$(ii)$] $\frac{1-p_i}{1-h_1\left(\mbox{\boldmath$p$}\right)}\frac{\partial h_1\left(\mbox{\boldmath$p$}\right)}{\partial p_i}$ or $\frac{1-p_i}{1-h_2\left(\mbox{\boldmath$p$}\right)}\frac{\partial h_2\left(\mbox{\boldmath$p$}\right)}{\partial p_i}$ is increasing in each $p_i$, $i=1,2,\dots,n$;
\item [$(iii)$] $X_i(\theta)\leq_{rhr}Y_i \text{ for all }\theta\in \Omega,$ and for all $i=1,2,\dots,n.$
\end{itemize}
Then $\tau_1\left(\mbox{\boldmath$X$}(\Theta)\right)\leq_{rhr}\tau_2\left(\mbox{\boldmath$Y$}\right)$.$\hfill\Box$
\end{thm}
\hspace*{0.3 in}Below we discuss  the corresponding result for the likelihood ratio order. See the Appendix for the proof.
\begin{thm}\label{t84}
Let $\mbox{\boldmath$X$}=(X_1,X_2,\dots,X_{n})$ and $\mbox{\boldmath$Y$}=(Y_1,Y_2,\dots,Y_{m})$ be two sets of components' lifetimes.
Suppose that the following conditions hold.
\begin{itemize}
 \item [$(i)$] For all $i=1,2,\dots,n \text{ and }j=1,2,\dots,m,$
  $$\frac{\frac{\partial h_2\left(\mbox{\boldmath$q$}\right)}{\partial q_j}}{\frac{\partial h_1\left(\mbox{\boldmath$p$}\right)}{\partial p_i}}\text{ is increasing in }x,\text{ for all }\theta\in \Omega,$$
  where $p_i=\bar F_{X_i}(x|\theta)$, $q_j=\bar F_{Y_j}(x)$;
\item [$(ii)$] $X_i(\theta)\leq_{lr}Y_j \text{ for all }\theta\in \Omega,$ and for all $i=1,2,\dots,n$ and $j=1,2,\dots,m.$
\end{itemize}
Then $\tau_1\left(\mbox{\boldmath$X$}(\Theta)\right)\leq_{lr}\tau_2\left(\mbox{\boldmath$Y$}\right)$.$\hfill\Box$
\end{thm}
\hspace*{0.2 in} This corollary follows from Theorem~\ref{t84} by using Lemma~\ref{le6}.
\begin{cor}
 Let $\mbox{\boldmath$X$}=(X_1,X_2,\dots,X_{n})$ and $\mbox{\boldmath$Y$}=(Y_1,Y_2,\dots,Y_{m})$ be two sets of component's lifetimes. Assume that the $X_i$'s are independent, and that the $Y_j$'s are independent.
 Suppose that condition ($ii$) in Theorem~\ref{t84} holds. Then
 \begin{itemize}
  \item [$(i)$] ${\tau_{k:n}\left(\mbox{\boldmath$X$}(\Theta)\right)}\leq_{lr}$ ${\tau_{l:n}\left(\mbox{\boldmath$Y$}\right)}$ for $l\leq k$;
  \item [$(ii)$] ${\tau_{k:n}\left(\mbox{\boldmath$X$}(\Theta)\right)}\leq_{lr}$ ${\tau_{k:m}\left(\mbox{\boldmath$Y$}\right)}$ for $n\leq m$;
  \item [$(iii)$] ${\tau_{k:n}\left(\mbox{\boldmath$X$}(\Theta)\right)}\leq_{lr}$ ${\tau_{k-r:n-r}\left(\mbox{\boldmath$Y$}\right)}$ for $r\leq k$;
  \item [$(iv)$] ${\tau_{k:n}\left(\mbox{\boldmath$X$}(\Theta)\right)}\leq_{lr}$ ${\tau_{l:m}\left(\mbox{\boldmath$Y$}\right)}$ for $l\leq k$ and $n-k\leq m-l$.
 \end{itemize}
\end{cor}
\subsection{Systems with iid components}
In this subsection, we consider coherent systems of iid components. In the following theorems, we compare
$\tau_1\left(\mbox{\boldmath$X$}(\Theta)\right)$ and $\tau_2\left(\mbox{\boldmath$Y$}\right)$ with respect to the usual stochastic order, the hazard rate order, the reversed hazard rate order and the likelihood ratio order. The proofs of these theorems could be done in the same line as in the previous subsection, and hence omitted.
\begin{thm}\label{t41}
Let $\mbox{\boldmath$X$}=(X_1,X_2,\dots,X_{n})$ and $\mbox{\boldmath$Y$}=(Y_1,Y_2,\dots,Y_{m})$ be two sets of components' lifetimes. Assume that $X_i$'s are identical, and that the $Y_j$'s are identical.
Suppose that the following conditions hold.
\begin{itemize}
\item [$(i)$] $h_1(p)\leq h_2(p)$ for all $p\in(0,1)$.
\item [$(ii)$] $X_1(\theta)\leq_{st}Y_1 \text{ for all }\theta\in \Omega,$.
\end{itemize}
Then $\tau_1\left(\mbox{\boldmath$X$}(\Theta)\right)\leq_{st}\tau_2\left(\mbox{\boldmath$Y$}\right)$.
\end{thm}
\begin{thm}\label{t42}
Let $\mbox{\boldmath$X$}=(X_1,X_2,\dots,X_{n})$ and $\mbox{\boldmath$Y$}=(Y_1,Y_2,\dots,Y_{m})$ be two sets of components' lifetimes. Assume that $X_i$'s are identical, and that the $Y_j$'s are identical.
Suppose that the following conditions hold.
\begin{itemize}
\item [$(i)$] $h_1(p)/ h_2(p)$ is increasing in $p\in (0,1)$;
\item [$(ii)$] $ph_1'(p)/h_1(p)$ or $ph_2'(p)/h_2(p)$ is decreasing in $p\in (0,1)$;
\item [$(ii)$] $X_1(\theta)\leq_{hr}Y_1  \text{ for all }\theta\in \Omega,$.
\end{itemize}
Then $\tau_1\left(\mbox{\boldmath$X$}(\Theta)\right)\leq_{hr}\tau_2\left(\mbox{\boldmath$Y$}\right)$.
\end{thm}
\begin{thm}\label{t43}
Let $\mbox{\boldmath$X$}=(X_1,X_2,\dots,X_{n})$ and $\mbox{\boldmath$Y$}=(Y_1,Y_2,\dots,Y_{m})$ be two sets of components' lifetimes. Assume that $X_i$'s are identical, and that the $Y_j$'s are identical.
Suppose that the following conditions hold.
\begin{itemize}
\item [$(i)$] $(1-h_1(p))/(1- h_2(p))$ is increasing in $p\in (0,1)$;
\item [$(ii)$] $(1-p)h_1'(p)/(1-h_1(p))$ or $(1-p)h_2'(p)/(1-h_2(p))$ is increasing in $p\in (0,1)$;
\item [$(ii)$] $X_1(\theta)\leq_{rhr}Y_1\text{ for all }\theta\in \Omega,$.
\end{itemize}
Then $\tau_1\left(\mbox{\boldmath$X$}(\Theta)\right)\leq_{rhr}\tau_2\left(\mbox{\boldmath$Y$}\right)$.
\end{thm}
\begin{thm}\label{t44}
Let $\mbox{\boldmath$X$}=(X_1,X_2,\dots,X_{n})$ and $\mbox{\boldmath$Y$}=(Y_1,Y_2,\dots,Y_{m})$ be two sets of components' lifetimes. Assume that $X_i$'s are identical, and that the $Y_j$'s are identical.
 Suppose that the following conditions holds.
\begin{itemize}
\item [$(i)$] $h_1'(p)/h_2'(p)$ is increasing in $p\in (0,1)$;
\item [$(ii)$] For $k=1$ or $2$, there exists some point $\mu \in (0,1)$ such that
 \begin{itemize}
   \item [$(a)$] $ph''_k(p)/h'_k(p)\;\text{is decreasing and positive for all}\;p \in(0,\mu),$ and
  \item [$(b)$] $(1-p)h''_k(p)/h'_k(p)\;\text{is decreasing and negative for all}\; p \in(\mu,1)$.
  \end{itemize}
\item [$(iii)$] $X_1(\theta)\leq_{lr}Y_1 \text{ for all }\theta\in \Omega,$.
\end{itemize}
Then $\tau_1\left(\mbox{\boldmath$X$}(\Theta)\right)\leq_{lr}\tau_2\left(\mbox{\boldmath$Y$}\right)$.
\end{thm}
\begin{r1}
It is worthwhile to mention, that a corollary corresponding to each theorem discussed in this subsection could be formulated similar to those given in Subsection~\ref{s51}.
\end{r1}
\section{Concluding Remarks}\label{se6}
 In this paper, we study an impact of a random environment on lifetimes of coherent systems with dependent lifetimes. There are two combined sources of the considered dependence. One results from  the dependence of the components of the coherent system operating in a deterministic environment and the other is due to the dependence of components of a system sharing the same random environment. We provide different sets of sufficient conditions for the corresponding stochastic comparisons and consider various scenarios, namely, ($i$) two coherent systems operate under the same random environment; ($ii$) two coherent systems  operate under two different random environments; ($iii$) one of the coherent systems operates under a random environment, whereas the other under a deterministic one. Furthermore, we show that some of the proposed results hold for the well known $k$-out-of-$n$ and $l$-out-of-$m$ systems. These systems are very common in practice.
\\\hspace*{0.2 in} Motivated by discussions in  Amini-Seresht et al.~\cite{azb}, we present solutions for some open problems formulated in this paper. We also generalize the results of these authors and present some new comparisons as well. Specifically, we provide different sets of sufficient conditions for one system to dominate the other one with respect to different stochastic orders, namely, usual stochastic order, hazard rate order, reversed hazard rate order and likelihood ratio order. Moreover, our methodology for obtaining comparisons also differs from that discussed in their paper.
\\\hspace*{0.2 in}Even though we have incorporated a large number of new results in this paper, there are still some open problems left behind. One of them is to generalize the results discussed in Sections~\ref{se3}~and~\ref{se4} to the likelihood ratio order.
\\\hspace*{0.2 in}We believe that the obtained results and the developed methodology can be helpful not only to the specialists in mathematical reliability theory but for design engineers, reliability analysts, etc., as engineering systems usually operate in different environments that are random and the proper comparisons of reliability characteristics can help in choosing or designing the most appropriate system (e.g., for performing a mission).

\subsection*{Acknowledgements}
\hspace*{0.2 in}The first author sincerely acknowledges the financial support from the IIITDM Kancheepuram, Chennai. The work of the second author was supported by National Research Foundation of South Africa (Grant no: 103613).
 
 {\bf Appendix}

{\bf Proof of Theorem~\ref{t00}:}
Consider the following two cases.
\\Case-I: Let $m\geq n$.
Note that
\begin{eqnarray*}
\bar F_{\tau_1\left(\mbox{\boldmath$X$}(\Theta)\right)}(x)&=&\int_\Omega h_1\left(\bar F_{X_1}(x|\theta),\bar F_{X_2}(x|\theta),\dots,\bar F_{X_n}(x|\theta)\right)dF_{\Theta}(\theta)
\\&\leq&\int_\Omega h_1\left(\bar F_{Y_1}(x|\theta),\bar F_{Y_2}(x|\theta),\dots,\bar F_{Y_n}(x|\theta)\right)dF_{\Theta}(\theta)
\\&\leq&\int_\Omega h_2\left(\bar F_{Y_1}(x|\theta),\bar F_{Y_2}(x|\theta),\dots,\bar F_{Y_m}(x|\theta)\right)dF_{\Theta}(\theta)
\\&=&\bar F_{\tau_2\left(\mbox{\boldmath$Y$}(\Theta)\right)}(x),
\end{eqnarray*}
where the first inequality follows from condition ($ii$) and the fact that $h_1\left(\mbox{\boldmath$p$}\right)$ is increasing in each $p_i$. The second inequality follows from condition ($ii$).
\\Case-II: Let $n> m$.
Note that
\begin{eqnarray*}
\bar F_{\tau_1\left(\mbox{\boldmath$X$}(\Theta)\right)}(x)&=&\int_\Omega h_1\left(\bar F_{X_1}(x|\theta),\bar F_{X_2}(x|\theta),\dots,\bar F_{X_n}(x|\theta)\right)dF_{\Theta}(\theta)
\\&\leq&\int_\Omega h_1\left(\bar F_{Y_1}(x|\theta),\bar F_{Y_2}(x|\theta),\dots,\bar F_{Y_m}(x|\theta),\bar F_{X_{m+1}}(x|\theta),\dots,\bar F_{X_{n}}(x|\theta)\right)dF_{\Theta}(\theta)
\\&\leq&\int_\Omega h_2\left(\bar F_{Y_1}(x|\theta),\bar F_{Y_2}(x|\theta),\dots,\bar F_{Y_m}(x|\theta)\right)dF_{\Theta}(\theta)
\\&=&\bar F_{\tau_2\left(\mbox{\boldmath$Y$}(\Theta)\right)}(x),
\end{eqnarray*}
where the first inequality follows from condition ($ii$) and the fact that $h_1\left(\mbox{\boldmath$p$}\right)$ is increasing in each $p_i$. The second inequality follows from condition ($ii$). Hence the result is proved.
$\hfill\Box$
 \\\\{\bf Proof of Theorem~\ref{th22}:} We only prove the result under the set of conditions $\{(i),(ii),(iii)\}$. The result could be proved in the same line under the second set of conditions.
Note that
\begin{eqnarray*}
\sum\limits_{i=1}^m r_{Y_i}(x|\theta_2)\left[\frac{p_i}{h_2\left(\mbox{\boldmath$p$}\right)}\frac{\partial h_2\left(\mbox{\boldmath$p$}\right)}{\partial p_i}\right]_{p_i=\bar F_{Y_i}(x|\theta_2)}
&\geq&\sum\limits_{i=1}^m r_{Y_i}(x|\theta_1)\left[\frac{p_i}{h_2\left(\mbox{\boldmath$p$}\right)}\frac{\partial h_2\left(\mbox{\boldmath$p$}\right)}{\partial p_i}\right]_{p_i=\bar F_{Y_i}(x|\theta_2)}
\\&\geq&\sum\limits_{i=1}^m r_{Y_i}(x|\theta_1)\left[\frac{q_i}{h_2\left(\mbox{\boldmath$q$}\right)}\frac{\partial h_2\left(\mbox{\boldmath$q$}\right)}{\partial p_i}\right]_{q_i=\bar F_{Y_i}(x|\theta_1)},
\end{eqnarray*}
where the first and the second inequalities follow from conditions ($ii$) and ($iii$). The above inequality is equivalent to the fact that
\begin{eqnarray*}
\frac{h_2\left(\bar F_{Y_1}(x|\theta_2),\bar F_{Y_2}(x|\theta_2),\dots,\bar F_{Y_m}(x|\theta_2)\right)}{h_2\left(\bar F_{Y_1}(x|\theta_1),\bar F_{Y_2}(x|\theta_1),\dots,\bar F_{Y_m}(x|\theta_1)\right)}\text{ is decreasing in }x>0,
\end{eqnarray*}
or equivalently,
\begin{eqnarray}
h_2\left(\bar F_{Y_1}(x|\theta),\bar F_{Y_2}(x|\theta),\dots,\bar F_{Y_m}(x|\theta)\right) \text{is RR}_2\text{ in }(x,\theta)\in(0,\infty)\times \Omega.\label{e85}
\end{eqnarray}
Further, we have
\begin{eqnarray*}
\sum\limits_{i=1}^n r_{X_i}(x|\theta)\left[\frac{p_i}{h_1\left(\mbox{\boldmath$p$}\right)}\frac{\partial h_1\left(\mbox{\boldmath$p$}\right)}{\partial p_i}\right]_{p_i=\bar F_{X_i}(x|\theta)}
&\geq&\sum\limits_{i=1}^m r_{X_i}(x|\theta)\left[\frac{p_i}{h_1\left(\mbox{\boldmath$p$}\right)}\frac{\partial h_1\left(\mbox{\boldmath$p$}\right)}{\partial p_i}\right]_{p_i=\bar F_{X_i}(x|\theta)}
\\&\geq&\sum\limits_{i=1}^m r_{Y_i}(x|\theta)\left[\frac{p_i}{h_2\left(\mbox{\boldmath$p$}\right)}\frac{\partial h_2\left(\mbox{\boldmath$p$}\right)}{\partial p_i}\right]_{p_i=\bar F_{X_i}(x|\theta)}
\\&\geq &\sum\limits_{i=1}^m r_{Y_i}(x|\theta)\left[\frac{q_i}{h_2\left(\mbox{\boldmath$q$}\right)}\frac{\partial h_2\left(\mbox{\boldmath$q$}\right)}{\partial q_i}\right]_{q_i=\bar F_{Y_i}(x|\theta)},
\end{eqnarray*}
where the first inequality holds because each term in the summation is nonnegative. The second inequality follows from conditions ($i$) and ($iii$), whereas the third inequality follows from ($ii$) and ($iii$). Thus the above expression can equivalently be written as
\begin{eqnarray}
\frac{h_2\left(\bar F_{Y_1}(x|\theta),\bar F_{Y_2}(x|\theta),\dots,\bar F_{Y_m}(x|\theta)\right)}{h_1\left(\bar F_{X_1}(x|\theta),\bar F_{X_2}(x|\theta),\dots,\bar F_{X_n}(x|\theta)\right)}\text{ is increasing in }x>0,\text{ for all }\theta \in \Omega.\label{e82}
\end{eqnarray}
Again, condition ($iii$) implies that, $\text{ for all } \ts_1\leq \ts_2, \text{ and for all }i=1,2,\dots,n\text{ and }j=1,2,\dots,m,$
\begin{eqnarray}\label{e83}
X_i(\theta_1)\leq_{st}X_i(\theta_2)\text{ and }Y_j(\theta_2)\leq_{st}Y_j(\theta_1).
\end{eqnarray}
On using this, we get
\begin{eqnarray*}
\frac{d}{d\theta}{h_1\left(\bar F_{X_1}(x|\theta),\bar F_{X_2}(x|\theta),\dots,\bar F_{X_n}(x|\theta)\right)}&=&\sum\limits_{i=1}^n\left[\frac{\partial h_1\left(\mbox{\boldmath$p$}\right)}{\partial p_i}\frac{dp_i}{d\theta}\right]_{p_i=\bar F_{X_i}(x|\theta)}\nonumber
\\&\geq &0,
\end{eqnarray*}
and
\begin{eqnarray*}
\frac{d}{d\theta}{h_2\left(\bar F_{Y_1}(x|\theta),\bar F_{Y_2}(x|\theta),\dots,\bar F_{Y_m}(x|\theta)\right)}&=&\sum\limits_{j=1}^m\left[\frac{\partial h_2\left(\mbox{\boldmath$q$}\right)}{\partial p_i}\frac{dp_i}{d\theta}\right]_{p_i=\bar F_{Y_i}(x|\theta)}\nonumber
\\&\leq &0,
\end{eqnarray*}
or equivalently,
\begin{eqnarray*}
\frac{1}{h_1\left(\bar F_{X_1}(x|\theta),\bar F_{X_2}(x|\theta),\dots,\bar F_{X_n}(x|\theta)\right)}\text{ is decreasing in }\theta \in \Omega,
\end{eqnarray*}
and
\begin{eqnarray*}
{h_2\left(\bar F_{Y_1}(x|\theta),\bar F_{Y_2}(x|\theta),\dots,\bar F_{Y_m}(x|\theta)\right)}\text{ is decreasing in }\theta \in \Omega.
\end{eqnarray*}
On combining these two get
\begin{eqnarray}
\frac{h_2\left(\bar F_{Y_1}(x|\theta),\bar F_{Y_2}(x|\theta),\dots,\bar F_{Y_m}(x|\theta)\right)}{h_1\left(\bar F_{X_1}(x|\theta),\bar F_{X_2}(x|\theta),\dots,\bar F_{X_n}(x|\theta)\right)}\text{ is decreasing in }\theta \in \Omega, \text{ for all } x>0.\label{e84}
\end{eqnarray}
On using \eqref{e85}, \eqref{e82} and \eqref{e84} in Lemma~\ref{le1}, we get, for $x_1\leq x_2,$
\begin{eqnarray*}
\frac{\int_\Omega h_1\left(\bar F_{X_1}(x_2|\theta),\bar F_{X_2}(x_2|\theta),\dots,\bar F_{X_n}(x_2|\theta)\right)dF_{\Theta}(\theta)}{\int_\Omega h_2\left(\bar F_{Y_1}(x_2|\theta),\bar F_{Y_2}(x_2|\theta),\dots,\bar F_{Y_m}(x_2|\theta)\right)dF_{\Theta}(\theta)}
\\\leq \frac{\int_\Omega h_1\left(\bar F_{X_1}(x_1|\theta),\bar F_{X_2}(x_1|\theta),\dots,\bar F_{X_n}(x_2|\theta)\right)dF_{\Theta}(\theta)}{\int_\Omega h_2\left(\bar F_{Y_1}(x_1|\theta),\bar F_{Y_2}(x_1|\theta),\dots,\bar F_{Y_m}(x_1|\theta)\right)dF_{\Theta}(\theta)},
\end{eqnarray*}
or equivalently,
\begin{eqnarray*}
\frac{\bar F_{\tau_1\left(\mbox{\boldmath$X$}(\Theta)\right)}(x)}{\bar F_{\tau_2\left(\mbox{\boldmath$Y$}(\Theta)\right)}(x)}=\frac{\int_\Omega h_1\left(\bar F_{X_1}(x|\theta),\bar F_{X_2}(x|\theta),\dots,\bar F_{X_n}(x|\theta)\right)dF_{\Theta}(\theta)}{\int_\Omega h_2\left(\bar F_{Y_1}(x|\theta),\bar F_{Y_2}(x|\theta),\dots,\bar F_{Y_m}(x|\theta)\right)dF_{\Theta}(\theta)}\text{ is decreasing in }x>0,
\end{eqnarray*} and hence $\tau_1\left(\mbox{\boldmath$X$}(\Theta)\right)\leq_{hr}\tau_2\left(\mbox{\boldmath$Y$}(\Theta)\right)$.$\hfill\Box$
\\\\ {\bf Proof of Theorem~\ref{th24}:} We only prove the result under the set of conditions $\{(i),(ii),(iii)\}$. The result could be proved in the same line under the second set of conditions.
Note that
\begin{eqnarray*}
\sum\limits_{i=1}^n \tilde r_{X_i}(x|\theta_2)\left[\frac{1-p_i}{1-h_1\left(\mbox{\boldmath$p$}\right)}\frac{\partial h_1\left(\mbox{\boldmath$p$}\right)}{\partial p_i}\right]_{p_i=\bar F_{X_i}(x|\theta_2)}
&\geq&\sum\limits_{i=1}^n \tilde r_{X_i}(x|\theta_1)\left[\frac{1-p_i}{1-h_1\left(\mbox{\boldmath$p$}\right)}\frac{\partial h_1\left(\mbox{\boldmath$p$}\right)}{\partial p_i}\right]_{p_i=\bar F_{X_i}(x|\theta_2)}
\\&\geq&\sum\limits_{i=1}^n \tilde r_{X_i}(x|\theta_1)\left[\frac{q_i}{1-h_1\left(\mbox{\boldmath$q$}\right)}\frac{\partial h_1\left(\mbox{\boldmath$q$}\right)}{\partial p_i}\right]_{q_i=\bar F_{X_i}(x|\theta_1)},
\end{eqnarray*}
where the first and the second inequalities follow from conditions ($ii$) and ($iii$). The above expression can equivalently be written as
\begin{eqnarray*}
\frac{1-h_1\left(\bar F_{X_1}(x|\theta_2),\bar F_{X_2}(x|\theta_2),\dots,\bar F_{X_n}(x|\theta_2)\right)}{1-h_1\left(\bar F_{X_1}(x|\theta_1),\bar F_{X_2}(x|\theta_1),\dots,\bar F_{X_n}(x|\theta_1)\right)}\text{ is increasing in }x>0,
\end{eqnarray*}
or equivalently,
\begin{eqnarray}
1-h_1\left(\bar F_{X_1}(x|\theta),\bar F_{X_2}(x|\theta),\dots,\bar F_{X_n}(x|\theta)\right) \text{is TP}_2\text{ in }(x,\theta)\in(0,\infty)\times \Omega.\label{e92}
\end{eqnarray}
Further, we have
\begin{eqnarray*}
\sum\limits_{i=1}^m \tilde r_{Y_i}(x|\theta)\left[\frac{1-p_i}{1-h_2\left(\mbox{\boldmath$p$}\right)}\frac{\partial h_2\left(\mbox{\boldmath$p$}\right)}{\partial p_i}\right]_{p_i=\bar F_{Y_i}(x|\theta)}
&\geq&\sum\limits_{i=1}^n \tilde r_{Y_i}(x|\theta)\left[\frac{1-p_i}{1-h_2\left(\mbox{\boldmath$p$}\right)}\frac{\partial h_2\left(\mbox{\boldmath$p$}\right)}{\partial p_i}\right]_{p_i=\bar F_{Y_i}(x|\theta)}
\\&\geq&\sum\limits_{i=1}^n \tilde r_{X_i}(x|\theta)\left[\frac{1-p_i}{1-h_1\left(\mbox{\boldmath$p$}\right)}\frac{\partial h_2\left(\mbox{\boldmath$p$}\right)}{\partial p_i}\right]_{p_i=\bar F_{Y_i}(x|\theta)}
\\&\geq &\sum\limits_{i=1}^n \tilde r_{X_i}(x|\theta)\left[\frac{1-q_i}{1-h_1\left(\mbox{\boldmath$q$}\right)}\frac{\partial h_1\left(\mbox{\boldmath$q$}\right)}{\partial q_i}\right]_{q_i=\bar F_{X_i}(x|\theta)},
\end{eqnarray*}
where the first inequality holds because each term in the summation is nonnegative. The second inequality follows from conditions ($i$) and ($iii$), whereas the third inequality follows from ($ii$) and ($iii$). Then the above expression can equivalently be written as
\begin{eqnarray}
\frac{1-h_2\left(\bar F_{Y_1}(x|\theta),\bar F_{Y_2}(x|\theta),\dots,\bar F_{Y_m}(x|\theta)\right)}{1-h_1\left(\bar F_{X_1}(x|\theta),\bar F_{X_2}(x|\theta),\dots,\bar F_{X_n}(x|\theta)\right)}\text{ is increasing in }x>0,\text{ for all }\theta \in \Omega.\label{e93}
\end{eqnarray}
Again, condition ($iii$) implies that, $\text{ for all } \ts_1\leq \ts_2, \text{ and for all }i=1,2,\dots,n\text{ and }j=1,2,\dots,m,$
\begin{eqnarray*}\label{e90}
X_i(\theta_1)\leq_{st}X_i(\theta_2)\text{ and }Y_j(\theta_2)\leq_{st}Y_j(\theta_1).
\end{eqnarray*}
On using this, we get
\begin{eqnarray*}
\frac{d}{d\theta}{\left[1-h_1\left(\bar F_{X_1}(x|\theta),\bar F_{X_2}(x|\theta),\dots,\bar F_{X_n}(x|\theta)\right)\right]}&=&-\sum\limits_{i=1}^n\left[\frac{\partial h_1\left(\mbox{\boldmath$p$}\right)}{\partial p_i}\frac{dp_i}{d\theta}\right]_{p_i=\bar F_{X_i}(x|\theta)}\nonumber
\\&\leq &0,
\end{eqnarray*}
and
\begin{eqnarray*}
\frac{d}{d\theta}{\left[1-h_2\left(\bar F_{Y_1}(x|\theta),\bar F_{Y_2}(x|\theta),\dots,\bar F_{Y_m}(x|\theta)\right)\right]}&=&-\sum\limits_{j=1}^m\left[\frac{\partial h_2\left(\mbox{\boldmath$q$}\right)}{\partial p_i}\frac{dp_i}{d\theta}\right]_{p_i=\bar F_{Y_i}(x|\theta)}\nonumber
\\&\geq &0,
\end{eqnarray*}
or equivalently,
\begin{eqnarray*}
{\frac{1}{1-h_1\left(\bar F_{X_1}(x|\theta),\bar F_{X_2}(x|\theta),\dots,\bar F_{X_n}(x|\theta)\right)}}\text{ is increasing in }\theta \in \Omega,
\end{eqnarray*}
and
\begin{eqnarray*}
{1-h_2\left(\bar F_{Y_1}(x|\theta),\bar F_{Y_2}(x|\theta),\dots,\bar F_{Y_m}(x|\theta)\right)}\text{ is increasing in }\theta \in \Omega.
\end{eqnarray*}
On combining, we get
\begin{eqnarray}
\frac{1-h_2\left(\bar F_{Y_1}(x|\theta),\bar F_{Y_2}(x|\theta),\dots,\bar F_{Y_m}(x|\theta)\right)}{1-h_1\left(\bar F_{X_1}(x|\theta),\bar F_{X_2}(x|\theta),\dots,\bar F_{X_n}(x|\theta)\right)}\text{ is increasing in }\theta \in \Omega, \text{ for all } x>0.\label{e94}
\end{eqnarray}
On using \eqref{e92}, \eqref{e93} and \eqref{e94} in Lemma~\ref{le1}, we get, for $x_1\leq x_2,$
\begin{eqnarray*}
\frac{\int_\Omega \left[1-h_2\left(\bar F_{Y_1}(x_2|\theta),\bar F_{Y_2}(x_2|\theta),\dots,\bar F_{Y_m}(x_2|\theta)\right)\right]dF_{\Theta}(\theta)}{\int_\Omega \left[1-h_1\left(\bar F_{X_1}(x_2|\theta),\bar F_{X_2}(x_2|\theta),\dots,\bar F_{X_n}(x_2|\theta)\right)\right]dF_{\Theta}(\theta)}
\\\geq \frac{\int_\Omega \left[1-h_2\left(\bar F_{Y_1}(x_1|\theta),\bar F_{Y_2}(x_1|\theta),\dots,\bar F_{Y_m}(x_1|\theta)\right)\right]dF_{\Theta}(\theta)}{\int_\Omega \left[1-h_1\left(\bar F_{X_1}(x_1|\theta),\bar F_{X_2}(x_1|\theta),\dots,\bar F_{X_n}(x_2|\theta)\right)\right]dF_{\Theta}(\theta)},
\end{eqnarray*}
or equivalently,
\begin{eqnarray*}
\frac{ F_{\tau_2\left(\mbox{\boldmath$Y$}(\Theta)\right)}(x)}{F_{\tau_1\left(\mbox{\boldmath$X$}(\Theta)\right)}(x)}=\frac{\int_\Omega \left[1-h_2\left(\bar F_{Y_1}(x|\theta),\bar F_{Y_2}(x|\theta),\dots,\bar F_{Y_m}(x|\theta)\right)\right]dF_{\Theta}(\theta)}{\int_\Omega \left[1-h_1\left(\bar F_{X_1}(x|\theta),\bar F_{X_2}(x|\theta),\dots,\bar F_{X_n}(x|\theta)\right)\right]dF_{\Theta}(\theta)}\text{ is increasing in }x>0,
\end{eqnarray*} and hence $\tau_1\left(\mbox{\boldmath$X$}(\Theta)\right)\leq_{rhr}\tau_2\left(\mbox{\boldmath$Y$}(\Theta)\right)$.$\hfill\Box$
\\\\ {\bf Proof of Theorem~\ref{t82}:}
We only prove the result under the condition $\{(i),(iii),(iv)\}$. The result follows similarly for the other case.
Now, from condition ($iv$), we have
\begin{eqnarray}
\min_{1\leq i\leq n}r_{X_i}(x|\theta)\geq \max_{1\leq i\leq m}r_{Y_i}(x)\text{ for all }\theta \in \Omega.\label{e881}
\end{eqnarray}
Then
\begin{eqnarray*}
&&\sum\limits_{i=1}^n r_{X_i}(x|\theta)\left[\frac{p_i}{h_1\left(\mbox{\boldmath$p$}\right)}\frac{\partial h_1\left(\mbox{\boldmath$p$}\right)}{\partial p_i}\right]_{p_i=\bar F_{X_i}(x|\theta)}\nonumber
\\&\geq &\min_{1\leq i\leq n}r_{X_i}(x|\theta)\sum\limits_{i=1}^n \left[\frac{p_i}{h_1\left(\mbox{\boldmath$p$}\right)}\frac{\partial h_1\left(\mbox{\boldmath$p$}\right)}{\partial p_i}\right]_{p_i=\bar F_{X_i}(x|\theta)}\nonumber
\\&\geq &\max_{1\leq i\leq m}r_{Y_i}(x)\sum\limits_{i=1}^m \left[\frac{p_i}{h_2\left(\mbox{\boldmath$p$}\right)}\frac{\partial h_2\left(\mbox{\boldmath$p$}\right)}{\partial p_i}\right]_{\substack{p_i=\bar F_{X_i}(x|\theta),\;i=1,2,\dots,\min\{m,n\},\\\left\{p_i=\bar F_{Y_i}(x),\;i=n+1,\dots,m\right\}I_{[m>n]}}}\nonumber
\\&\geq &\max_{1\leq i\leq m}r_{Y_i}(x)\sum\limits_{i=1}^m\nonumber \left[\frac{q_i}{h_2\left(\mbox{\boldmath$q$}\right)}\frac{\partial h_2\left(\mbox{\boldmath$q$}\right)}{\partial q_i}\right]_{q_i=\bar F_{Y_i}(x)}\nonumber
\\&\geq &\sum\limits_{i=1}^m r_{Y_i}(x)\left[\frac{q_i}{h_2\left(\mbox{\boldmath$q$}\right)}\frac{\partial h_2\left(\mbox{\boldmath$q$}\right)}{\partial q_i}\right]_{q_i=\bar F_{Y_i}(x)},\label{885}
\end{eqnarray*}
where the first and the fourth inequalities are obvious. The second inequality follows from \eqref{e881} and condition ($i$), whereas the third inequality follows from condition ($iii$) and ($iv$). Now, the above inequality implies that
\begin{eqnarray*}
 h_2\left(\bar F_{Y_1}(x),\bar F_{Y_2}(x),\dots,\bar F_{Y_m}(x)\right)\int_\Omega \left(\sum\limits_{i=1}^n r_{X_i}(x|\theta)\left[{p_i}\frac{\partial h_1\left(\mbox{\boldmath$p$}\right)}{\partial p_i}\right]_{p_i=\bar F_{X_i}(x|\theta)}\right)dF_{\Theta}(\theta)\nonumber
\\\geq \left(\sum\limits_{i=1}^m r_{Y_i}(x)\left[{q_i}\frac{\partial h_2\left(\mbox{\boldmath$q$}\right)}{\partial q_i}\right]_{q_i=\bar F_{Y_i}(x)}\right)\int_\Omega h_1\left(\bar F_{X_1}(x|\theta),\bar F_{X_2}(x|\theta),\dots,\bar F_{X_n}(x|\theta)\right)dF_{\Theta}(\theta),\label{eq880}
\end{eqnarray*}
which is equivalent to the fact that
\begin{eqnarray*}
\frac{\bar F_{\tau_1\left(\mbox{\boldmath$X$}(\Theta)\right)}(x)}{\bar F_{\tau_2\left(\mbox{\boldmath$Y$}\right)}(x)}=\frac{\int_\Omega h_1\left(\bar F_{X_1}(x|\theta),\bar F_{X_2}(x|\theta),\dots,\bar F_{X_n}(x|\theta)\right)dF_{\Theta}(\theta)}{ h_2\left(\bar F_{Y_1}(x),\bar F_{Y_2}(x),\dots,\bar F_{Y_m}(x)\right)}\text{ is decreasing in }x>0,
\end{eqnarray*}
 and hence $\tau_1\left(\mbox{\boldmath$X$}(\Theta)\right)\leq_{hr}\tau_2\left(\mbox{\boldmath$Y$}\right)$.$\hfill\Box$
 \\\\ {\bf Proof of Theorem~\ref{t83}:} We only prove the result under the condition $\{(i),(ii),(iv)\}$. The result follows similarly for the other case.
Now, from condition ($iv$), we have
\begin{eqnarray}
\min_{1\leq i\leq m}\tilde r_{Y_i}(x)\geq \max_{1\leq i\leq n}\tilde r_{X_i}(x|\theta)\text{ for all }\theta \in \Omega.\label{e887}
\end{eqnarray}
Note that
\begin{eqnarray*}
&&\sum\limits_{i=1}^m \tilde r_{Y_i}(x)\left[\frac{1-p_i}{1-h_2\left(\mbox{\boldmath$p$}\right)}\frac{\partial h_2\left(\mbox{\boldmath$p$}\right)}{\partial p_i}\right]_{p_i=\bar F_{Y_i}(x)}
\\&\geq &\min_{1\leq i\leq m}\tilde r_{Y_i}(x)\sum\limits_{i=1}^m \left[\frac{1-p_i}{1-h_2\left(\mbox{\boldmath$p$}\right)}\frac{\partial h_2\left(\mbox{\boldmath$p$}\right)}{\partial p_i}\right]_{p_i=\bar F_{Y_i}(x)}
\\&\geq &\max_{1\leq i\leq n}\tilde r_{X_i}(x|\theta)\sum\limits_{i=1}^n \left[\frac{1-p_i}{1-h_1\left(\mbox{\boldmath$p$}\right)}\frac{\partial h_1\left(\mbox{\boldmath$p$}\right)}{\partial p_i}\right]_{\substack{p_i=\bar F_{Y_i}(x),\;i=1,2,\dots,\min\{m,n\},\\\left\{p_i=\bar F_{X_i}(x|\theta),\;i=m+1,\dots,n\right\}I_{[n>m]}}}
\\&\geq &\max_{1\leq i\leq n}\tilde r_{X_i}(x|\theta)\sum\limits_{i=1}^n \left[\frac{1-q_i}{1-h_1\left(\mbox{\boldmath$q$}\right)}\frac{\partial h_1\left(\mbox{\boldmath$q$}\right)}{\partial q_i}\right]_{q_i=\bar F_{X_i}(x|\theta)}
\\&\geq &\sum\limits_{i=1}^n \tilde r_{X_i}(x|\theta)\left[\frac{1-q_i}{1-h_1\left(\mbox{\boldmath$q$}\right)}\frac{\partial h_1\left(\mbox{\boldmath$q$}\right)}{\partial q_i}\right]_{q_i=\bar F_{X_i}(x|\theta)},
\end{eqnarray*}
where the first and the fourth inequalities are obvious. The second inequality follows from \eqref{e887} and condition ($i$), whereas the third inequality follows from condition ($ii$) and ($iv$). Further, the above inequality implies that
\begin{eqnarray*}
 \left(\sum\limits_{i=1}^m \tilde r_{Y_i}(x)\left[{(1-p_i)}\frac{\partial h_2\left(\mbox{\boldmath$p$}\right)}{\partial p_i}\right]_{p_i=\bar F_{Y_i}(x)}\right)\int_\Omega\left[1- h_1\left(\bar F_{X_1}(x|\theta),\bar F_{X_2}(x|\theta),\dots,\bar F_{X_n}(x|\theta)\right)\right]dF_{\Theta}(\theta)\nonumber
\\\geq \left[1- h_2\left(\bar F_{Y_1}(x),\bar F_{Y_2}(x),\dots,\bar F_{Y_m}(x)\right)\right]\int_\Omega\left(\sum\limits_{i=1}^n \tilde r_{X_i}(x|\theta)\left[{(1-q_i)}\frac{\partial h_1\left(\mbox{\boldmath$q$}\right)}{\partial q_i}\right]_{q_i=\bar F_{X_i}(x|\theta)}\right)dF_{\Theta}(\theta),
\end{eqnarray*}
which is equivalent to the fact that
\begin{eqnarray*}
\frac{ F_{\tau_1\left(\mbox{\boldmath$X$}(\Theta)\right)}(x)}{ F_{\tau_2\left(\mbox{\boldmath$Y$}\right)}(x)}=\frac{\int_\Omega\left[1- h_1\left(\bar F_{X_1}(x|\theta),\bar F_{X_2}(x|\theta),\dots,\bar F_{X_n}(x|\theta)\right)\right]dF_{\Theta}(\theta)}{1- h_2\left(\bar F_{Y_1}(x),\bar F_{Y_2}(x),\dots,\bar F_{Y_m}(x)\right)}\text{ is decreasing in }x>0,
\end{eqnarray*}
 and hence $\tau_1\left(\mbox{\boldmath$X$}(\Theta)\right)\leq_{rhr}\tau_2\left(\mbox{\boldmath$Y$}\right)$.$\hfill\Box$
 \\\\ {\bf Proof of Theorem~\ref{t84}:} Note that
 $\tau_1\left(\mbox{\boldmath$X$}(\Theta)\right)\leq_{lr}\tau_2\left(\mbox{\boldmath$Y$}\right)$ holds if
 \begin{eqnarray*}
 \frac{ f_{\tau_1\left(\mbox{\boldmath$X$}(\Theta)\right)}(x)}{ f_{\tau_2\left(\mbox{\boldmath$Y$}\right)}(x)}= \frac{\int_\Omega \left[\sum\limits_{i=1}^n \left(f_{X_i}(x|\theta)\frac{\partial h_1\left(\mbox{\boldmath$p$}\right)}{\partial p_i} \right)\right]dF_{\Theta}(\theta)}{\sum\limits_{i=1}^m \left(f_{Y_i}(x)\frac{\partial h_2\left(\mbox{\boldmath$q$}\right)}{\partial q_i}\right)} \text{ is decreasing in }x>0,
 \end{eqnarray*}
 or equivalently,
 \begin{eqnarray*}
 \frac{\sum\limits_{i=1}^n \left[ \int_\Omega\left(f_{X_i}(x|\theta)\frac{\partial h_1\left(\mbox{\boldmath$p$}\right)}{\partial p_i} \right)dF_{\Theta}(\theta)\right]}{\sum\limits_{i=1}^m \left(f_{Y_i}(x)\frac{\partial h_2\left(\mbox{\boldmath$q$}\right)}{\partial q_i}\right)} \text{ is decreasing in }x>0.
 \end{eqnarray*}
 This holds if, for all $i=1,2,\dots,n$ and $j=1,2,\dots,m$,
 \begin{eqnarray*}
\int_\Omega\left(\frac{f_{X_i}(x|\theta)}{f_{Y_j}(x)}\right)\left(\frac{\frac{\partial h_1\left(\mbox{\boldmath$p$}\right)}{\partial p_i}}{\frac{\partial h_2\left(\mbox{\boldmath$q$}\right)}{\partial q_j}}\right)dF_{\Theta}(\theta) \text{ is decreasing in }x>0,
 \end{eqnarray*}
 which holds if, for all $i=1,2,\dots,n$ and $j=1,2,\dots,m$,
 \begin{eqnarray*}
 \frac{f_{X_i}(x|\theta)}{f_{Y_j}(x)} \text{ is decreasing in }x>0, \text{ for all } \theta \in \Omega,\label{e01}
 \end{eqnarray*}
and
 \begin{eqnarray*}\label{e04}
\frac{\frac{\partial h_1\left(\mbox{\boldmath$p$}\right)}{\partial p_i}}{\frac{\partial h_2\left(\mbox{\boldmath$q$}\right)}{\partial q_j}} \text{ is decreasing in }x>0,
 \end{eqnarray*}
 which are true because of conditions ($i$) and ($ii$).$\hfill\Box$
\end{document}